\documentclass[12pt]{article}

\usepackage[bulletsep]{collref}
\usepackage{graphicx}
\usepackage{pst-all}
\usepackage{bbm}
\usepackage{amsmath}
\usepackage{amssymb}
\usepackage{subfigure}
\usepackage{slashed}
\usepackage{psfrag}
\usepackage{array}
\usepackage{verbatim}
\usepackage{feynmp}

\usepackage{diagrams}
\diagramstyle[shortfall=4mm]

\makeatletter \@addtoreset{equation}{section} \makeatother

\makeatletter
\let\old@startsection=\@startsection
\let\oldl@section=\l@section
\renewcommand{\@startsection}[6]{\old@startsection{#1}{#2}{#3}{#4}{#5}{#6\mathversion{bold}}}
\renewcommand{\l@section}[2]{\oldl@section{\mathversion{bold}#1}{#2}}
\makeatother

\makeatletter
\let\old@makecaption=\@makecaption
\def\@makecaption{\small\old@makecaption}
\makeatother

\renewcommand{\thefootnote}{\arabic{footnote}}
\let\oldPhi=\Phi
\let\oldPsi=\Psi
\let\oldGamma=\Gamma
\let\oldDelta=\Delta
\let\oldSigma=\Sigma
\let\oldTheta=\Theta
\let\oldPi=\Pi
\let\oldUpsilon=\Upsilon
\renewcommand{\Phi}{\mathnormal{\oldPhi}}
\renewcommand{\Psi}{\mathnormal{\oldPsi}}
\renewcommand{\Gamma}{\mathnormal{\oldGamma}}
\renewcommand{\Sigma}{\mathnormal{\oldSigma}}
\renewcommand{\Delta}{\mathnormal{\oldDelta}}
\renewcommand{\Theta}{\mathnormal{\oldTheta}}
\renewcommand{\Pi}{\mathnormal{\oldPi}}
\renewcommand{\Upsilon}{\mathnormal{\oldUpsilon}}

\newcommand{\superN}{\mathcal{N}}

\newcommand{\Lagr}{\mathcal{L}}

\newcommand{\diag}{\mathop{\mathrm{diag}}}

\newcommand{\order}{\mathcal{O}}

\newcommand{\trans}{{\scriptscriptstyle\mathrm{T}}}

\ifx\genfrac\sdflkaj

\else

\fi
\newcommand{\sfrac}[2]{{\textstyle\frac{#1}{#2}}}
\newcommand{\half}{\sfrac{1}{2}}

\newcommand{\quarter}{\sfrac{1}{4}}

\newcommand{\iHalf}{\frac{i}{2}}
\newcommand{\Quarter}{\frac{1}{4}}
\newcommand{\p}{\partial}

\newcommand{\matr}[2]{\left(\begin{array}{#1}#2\end{array}\right)}

\newcommand{\grp}[1]{\mathrm{#1}}

\newcommand{\grU}{\grp{U}}
\newcommand{\grSU}{\grp{SU}}

\newcommand{\lrbrk}[1]{\left(#1\right)}
\newcommand{\bigbrk}[1]{\bigl(#1\bigr)}
\newcommand{\Bigbrk}[1]{\Bigl(#1\Bigr)}
\newcommand{\biggbrk}[1]{\biggl(#1\biggr)}

\newcommand{\lrsbrk}[1]{\left[#1\right]}
\newcommand{\bigsbrk}[1]{\bigl[#1\bigr]}
\newcommand{\Bigsbrk}[1]{\Bigl[#1\Bigr]}
\newcommand{\biggsbrk}[1]{\biggl[#1\biggr]}
\newcommand{\Biggsbrk}[1]{\Biggl[#1\Biggr]}
\newcommand{\ket}[1]{\mathopen{|}#1\mathclose{\rangle}}
\newcommand{\bra}[1]{\mathopen{\langle}#1\mathclose{|}}
\newcommand{\braket}[2]{\mathopen{\langle}#1|#2\mathclose{\rangle}}
\newcommand{\vev}[1]{\langle#1\rangle}

\newcommand{\biggvev}[1]{\biggl\langle#1\biggr\rangle}

\newcommand{\comm}[2]{[#1,#2]}

\newcommand{\acomm}[2]{\{#1,#2\}}

\newcommand{\abs}[1]{{|#1|}}

\newcommand{\bigeval}[1]{#1\big|}

\newcommand{\nn}{\nonumber}
\newcommand{\nln}{\nonumber\\}
\newcommand{\nl}[1][0pt]{\nonumber\\[#1]&\hspace{-4\arraycolsep}&\mathord{}}

\newcommand{\earel}[1]{\mathrel{}&\hspace{-2\arraycolsep}#1\hspace{-2\arraycolsep}&\mathrel{}}
\newcommand{\eq}{\earel{=}}

\def\[{\begin{equation}}
\def\]{\end{equation}}

\makeatletter
\def\mr@ignsp#1 {\ifx\:#1\@empty\else #1\expandafter\mr@ignsp\fi}%
\newcommand{\multiref}[1]{\begingroup
\xdef\mr@no@sparg{\expandafter\mr@ignsp#1 \: }%
\def\mr@comma{}%
\@for\mr@refs:=\mr@no@sparg\do{\mr@comma\def\mr@comma{,}\ref{\mr@refs}}%
\endgroup}
\makeatother

\newcommand{\hypref}[2]{\ifx\href\asklfhas #2\else\href{#1}{#2}\fi}

\newcommand{\secref}[1]{Sec.~\multiref{#1}}

\newcommand{\appref}[1]{App.~\multiref{#1}}

\newcommand{\figref}[1]{Fig.~\multiref{#1}}
\renewcommand{\eqref}[1]{(\multiref{#1})}

\ifx\href\asklfhas\newcommand{\href}[2]{#2}\fi

\newcommand{\comma}{\quad,\quad}
\newcommand{\unit}{\mathbbm{1}}

\newcommand{\levi}{\epsilon}

\newcommand{\eps}{\varepsilon}

\newcommand{\be}{\begin{eqnarray}}
\newcommand{\ee}{\end{eqnarray}}

\newcommand{\auxD}{\mathsf{D}}
\newcommand{\deriD}{\mathcal{D}}

\newcommand{\VV}{\mathcal{V}}

\newcommand{\dalpha}{{\dot{\alpha}}}
\newcommand{\dbeta}{{\dot{\beta}}}

\def\q{{{\vec q}}}
\def\p{{{\vec p}}}

\begin{document}

\thispagestyle{empty}
\begin{flushright}\footnotesize
\texttt{arXiv:0912.0733}\\
\texttt{PUPT-2327}\vspace{10mm}
\end{flushright}

\renewcommand{\thefootnote}{\fnsymbol{footnote}}
\setcounter{footnote}{0}

\begin{center}
{\Large\textbf{\mathversion{bold}
The Perfect Atom: \\
{\large Bound States of Supersymmetric Quantum Electrodynamics}
}\par}

\vspace{1.5cm}

\textrm{Christopher P.\ Herzog$^{a}$ and Thomas Klose$^{b}$} \vspace{8mm} \\
\textit{
$^a$Joseph Henry Laboratories and $^b$Princeton Center for Theoretical Science \\
Princeton University, Princeton, NJ 08544, USA
} \\
\texttt{\\ cpherzog,tklose@princeton.edu}

\par\vspace{14mm}

\textbf{Abstract} \vspace{5mm}

\begin{minipage}{14cm}
We study hydrogen-like atoms in $\superN=1$ supersymmetric quantum electrodynamics with an electronic and a muonic family. These atoms are bound states of an anti-muon and an electron or their superpartners. The exchange of a photino converts different bound states into each other. We determine the energy eigenstates and calculate the spectrum to fourth order in the fine structure constant. A difference between these perfect atoms and non-supersymmetric ones is the absence of hyperfine structure. We organize the eigenstates into super multiplets of the underlying symmetry algebra.
\end{minipage}

\end{center}

\vspace{0.5cm}

\break
\setcounter{page}{1}
\renewcommand{\thefootnote}{\arabic{footnote}}
\setcounter{footnote}{0}

\hrule
\tableofcontents
\vspace{8mm}
\hrule
\vspace{4mm}

\newpage
\section{Introduction}

Supersymmetry is often invoked to resolve a number of theoretical difficulties with the standard model of particle physics.  The symmetry can control quantum corrections to the Higgs mass, thus providing a solution to the hierarchy problem.  The symmetry suggests the strong force, the weak force, and electrogmagnetism are unified at high energy scales.  Moreover, the lightest supersymmetric partner is a candidate for dark matter. Despite these theoretical advantages, we see no direct evidence for supersymmetry at low energies; supersymmetry must be broken, and most studies of supersymmetry are devoted to investigating methods for and consequences of the breaking.
In this paper, we take a different tack and look at the energy spectrum of an anti-muon electron bound state in a theory with unbroken supersymmetry, supersymmetric quantum electrodynamics (SQED).

We were initially inspired to write this paper by work on gauge/gravity duality.  A gauge/gravity duality is a map between a field theory and a string theory.  The duality is useful because when the field theory is strongly interacting, the string theory is weakly interacting and vice versa.  Both the field theory and the string theory are typically supersymmetric.  Often one is faced with the following awkward situation: A calculation on the gravity side has revealed some property of the strongly interacting field theory, and the corresponding property of the field theory at weak coupling has not yet been studied.  One prime example of such a situation was the computation of the viscosity of $\superN=4$ $\grSU(N)$ super Yang-Mills theory at strong coupling in ref.\ \cite{Policastro:2001yc}.  Only five years later was the viscosity calculated in the perturbative limit \cite{Huot:2006ys}. In the case of supersymmetric atoms, ref.\ \cite{Paredes:2004is,Erdmenger:2006bg,Erdmenger:2007vj,Herzog:2008bp} studied hydrogenic bound states at strong coupling in $\superN=4$ $\grSU(N)$ super Yang-Mills modified by the addition of two massive $\superN=2$ hypermultiplets.\footnote{Given the strongly interacting nature of the bound states, heavy-light or hybrid meson is perhaps more appropriate terminology.} No corresponding study at the time had been made of such bound states at weak coupling. Moreover, the interesting observation was made that these bound states exhibited no hyperfine structure \cite{Herzog:2008bp}. The absence of such structure is an almost trivial consequence of $\superN=2$ supersymmetry \cite{DiVecchia:1985xm}, but in this paper we will see that hydrogenic atoms of $\superN=1$ SQED also lack hyperfine structure. It should, however, be emphasized that the energy spectra of $\superN=1$ and $\superN=2$ SQED remain noticeably different. The energy levels of $\superN=2$ hydrogenic atoms are independent of the spin of both the electron and the proton, while fine structure effects remain evident for $\superN=1$ atoms.


A second motivation for this paper is pure intellectual curiosity.  Although the 1s state of supersymmetric positronium was considered almost thirty years ago \cite{Buchmuller:1981bp}, no one to our knowledge has studied anti-muon electron bound states in SQED.  The way in which the bound states organize themselves into supermultiplets is surprising and intricate.  Similar to what happens for supersymmetric positronium, both degenerate and second order perturbation theory contribute at the same order in the fine structure constant $\alpha$.

We hope that these super atoms may be useful in particle physics, perhaps as a candidate for dark matter, perhaps in a hidden sector, perhaps for neutrino physics.

The paper is organized as follows.  In section \ref{sec:SQED}, we review SQED.  In section \ref{sec:As}, we present the relevant scattering amplitudes necessary for computing the energy spectrum to order $\alpha^4$.  In section \ref{sec:mixing}, we reduce the energy spectrum computation from field theory to time independent perturbation theory in quantum mechanics.  Section \ref{sec:eigenstates} contains detailed results for the hydrogenic states and their energies.
Section \ref{sec:conclusions} contains some discussion.
First, however, we summarize our results.

\subsection{Results}
\label{sec:results}

The energy spectrum of the hydrogen atom is usually described order by order in the fine structure constant $\alpha$.  The rest mass of the atom is $M+m$ where $M$ is the mass of the proton and $m$ the mass of the electron.  The binding energy is of order $\alpha^2 \mu$ where $\mu = M m / (M+m)$ is the reduced mass and we work in units where the speed of light $c=1$.  Fine structure effects are of order $\alpha^4 \mu$ and involve relativistic corrections along with spin-orbit couplings of the electron's spin to its orbital angular momentum. Hyperfine structure is of order $\alpha^4 \mu m / M$ and involves spin-spin coupling of the electron and proton.  
There are higher order corrections, for example the Lamb shift at order $\alpha^5$, but in this paper we work only to order $\alpha^4$.  

The proton is a composite object in the real world, and its compositeness has subtle effects on the hydrogen spectrum that do not interest us for the purposes of this paper.  Thus, we replace the proton with a fundamental particle of positive charge, an anti-muon.  Although in the case of the hydrogen atom $M$ is much larger than $m$, the results we present are valid for arbitrary values of $m$ and $M$.

We consider SQED in 3+1 dimensions with four super charges.  The electron and muon have super partners, the selectron and smuon.  Because the electron and muon have both charge and a Dirac mass, they need to be Dirac fermions and as such will each have two complex scalar field super partners.  In other words, there are {\it two} selectrons and {\it two} smuons.  The super partner of the photon is a Majorana spinor, the photino.

The existence of these super partners leads to some interesting effects.  An electron anti-muon bound state can mutate into a selectron anti-smuon bound state and back through photino exchange.  There are also fermionic bound states: an electron anti-smuon or a selectron anti-muon which can mutate into each other.  Because of photino exchange, the eigenstates of the effective Hamiltonian describing our super atom are actually linear superpositions of these different types of bound states.

The total angular momentum is a good quantum number for the bound states and is a useful organizing principle for the energy spectrum.
Consider a hydrogenic wave function with principal quantum number $n$ and orbital angular momentum $l>0$.  Let $V_l$ be a $2l+1$ dimensional representation of the $SO(3)$ rotation group.  A bound state of an electron and an anti-muon will transform as 
\be
V_l \otimes V_{1/2} \otimes V_{1/2} = V_{l-1} \oplus 2 V_l \oplus V_{l+1}
\ee
under $SO(3)$.  As there are two selectrons and two smuons, there are four fermionic bound states consisting either of an electron anti-smuon or selectron anti-muon.  These bound states transform as 
\be
4 (V_l \otimes V_{1/2}) = 4 V_{l-1/2} \oplus 4 V_{l+1/2} \; .
\ee
Finally, there are four bosonic bound states consisting of a selectron and anti-smuon, all transforming as $V_{l}$.  

Given supersymmetry, the energy spectrum must organize itself into super multiplets arising from the four super charges in 3+1 dimensions. If $j$ is a total angular momentum quantum number, then for $j>0$, a massive super multiplet consists of the four representations of the rotation group ${\mathcal R}_j = V_{j-1/2} \oplus 2 V_j \oplus V_{j+1/2}$. From the analysis in the previous paragraph, we see that to each hydrogenic wave function of principle quantum number $n$ and orbital angular momentum $l>0$, we can associate the four super multiplets
\begin{equation}
\label{multiplettable}
\begin{array}{c|cccc}
j  & \mathcal{R}_{l-1/2} & \mathcal{R}_{l} & \mathcal{R}_{l} & \mathcal{R}_{l+1/2} \\
\hline
l-1 & 1 \\
l-\frac{1}{2} & 2 & 1 & 1 \\
l & 1 & 2 & 2 & 1 \\
l+\frac{1}{2} & &1 & 1 & 2 \\
l+1 & & & & 1
\end{array}
\end{equation}

\begin{figure}
\begin{center}
\includegraphics[scale=1]{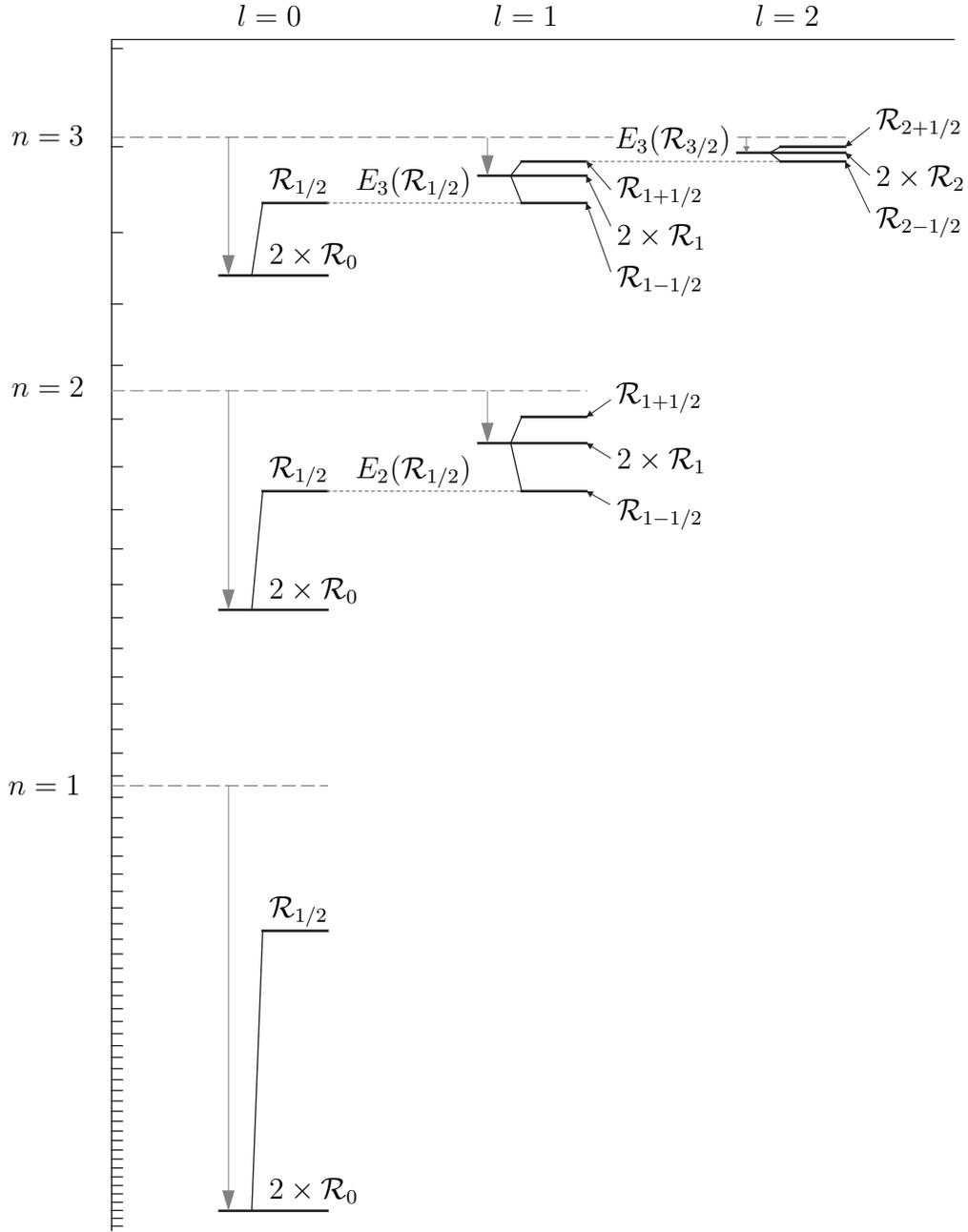}
\caption{\textbf{Spectrum.} Splitting of the energy levels $n=1,2,3$ with angular momentum $l=0,1,2$. The labels on the lines indicate the super multiplets. The bare energies $-\frac{\mu\alpha^2}{2n^2}$ are shifted due to a relativistic correction to the kinetic energy by 
$\delta E_{nl}$ as given in \protect\eqref{eqn:energyshift}, and then split due to various interactions into two or three levels.}
\label{fig:spectrum}
\end{center}
\end{figure}

Our results for the first few atomic energy levels of supersymmetric hydrogen are given in Figure \ref{fig:spectrum}. More generally, we find that the two multiplets ${\mathcal R}_l$ are degenerate in energy. The multiplet ${\mathcal R}_{l+1/2}$ is higher in energy by an amount
\be
\Delta E = \frac{ \mu \alpha^4}{2 (l+1)(2l+1) n^3} \ ,
\label{plusE}
\ee
while the multiplet ${\mathcal R}_{l-1/2}$ is lower in energy by an amount
\be
\Delta E = -\frac{\mu \alpha^4}{2l (2l+1) n^3} \ .
\label{minusE}
\ee
In the case $j=0$, the ${\mathcal R}_{-1/2}$ multiplet of course does not exist and the ${\mathcal R}_0$ multiplets do not contain the representation $V_{-1/2}$:
\begin{equation}
\label{multiplettablesmall}
\begin{array}{c|ccc}
j & \mathcal{R}_{0} & \mathcal{R}_{0} & \mathcal{R}_{1/2} \\
\hline
0  & 2 & 2 & 1 \\
\frac{1}{2} & 1 & 1 & 2 \\
1 &  & & 1
\end{array}
\end{equation}
The two ${\mathcal R}_0$ multiplets are degenerate in energy, and the ${\mathcal R}_{1/2}$ multiplet is higher in energy by an amount $\mu \alpha^4 / 2 n^3$. 

There is also an overall shift in the energies at order $\alpha^4$, see \eqref{eqn:energyshift}. For the total energy of a state in a super multiplet ${\mathcal R}_j$, we find
\be
E_n({\mathcal R}_j) = - \frac{\mu \alpha^2}{2 n^2} - \frac{\mu \alpha^4}{n^4}
\left( \frac{n}{2j+1} - \frac{3}{8} + \frac{\mu^2}{8Mm} \right)  + {\mathcal O}(\alpha^5)\ .
\label{finalanswer}
\ee
As is also true in QED, this expression can be written purely in terms of $j$.  In other words, the super multiplet ${\mathcal R}_{l+1/2}$ that comes from an $\ket{nl}$ state and the ${\mathcal R}_{(l+1)-1/2}$ super multiplet that comes from an $\ket{n,l+1}$ state are degenerate in energy.

An important observation about these energy splittings \eqref{plusE} and \eqref{minusE} is the absence of hyperfine structure; there is no energy 
splitting between two states of order $\alpha^4 \mu m / M$ in the case $m \ll M$. This absence is an effect of supersymmetry. In pure QED, we find the multiplets $V_{l}\oplus V_{l+1} \equiv {\mathcal A}_{l+1/2}$ ($l=0,1,2,...$) and $V_{l-1}\oplus V_{l} \equiv {\mathcal A}_{l-1/2}$ ($l=1,2,...$). In the limit $m \ll M$, there is a fine structure splitting between ${\mathcal A}_{l+1/2}$ and ${\mathcal A}_{l-1/2}$, and a further hyperfine splitting between the $V_l$ and $V_{l \pm 1}$ in ${\mathcal A}_{l\pm 1/2}$.

In cases with more supersymmetry, the corresponding multiplets are even larger and the splitting \eqref{plusE} and \eqref{minusE} will disappear as well \cite{DiVecchia:1985xm}. For example in a theory with 8 supercharges, a massive multiplet with $j\geq 1$ transforms as $V_{j-1} \oplus 4 V_{j-1/2} \oplus 6 V_j \oplus 4 V_{j+1/2} \oplus V_{j+1}$ under the rotation group. To get such a large multiplet, we need to combine the four multiplets in table \eqref{multiplettable}. Similarly, in the case $j=0$, we would need to combine the three multiplets in table \eqref{multiplettablesmall}.

It is also instructive to compare our results for muonium to the splitting of the ground state energy of $\superN=1$ positronium found in \cite{Buchmuller:1981bp}. The two computations differ in the respect that the latter one involves two particles of the same mass that moreover can annihilate. Setting $M=m$ and $n=1$ in \eqref{finalanswer}, we find that the two levels at $l=0$ differ in energy by $\Delta E = \frac{m\alpha^4}{4}$ which is half the value for the splitting between the ortho and para states of positronium \cite{Buchmuller:1981bp}. This difference is a consequence of the absence of the annihilation diagrams.

\section{SQED}
\label{sec:SQED}

We write $\superN=1$ SQED for two families of matter particles which we refer to as ``electronic'' and ``muonic.'' The electron $e^-$, its superpartners the selectrons $\tilde{e}_\pm^-$, and their antiparticles $e^+$ and $\tilde{e}_\pm^+$, are collectively represented by two chiral superfields $\Phi_{e\pm}$ with $\grU(1)$ charge $\pm e$ and mass $m$. Similarly, we write $\Phi_{m\pm}$ for the muon $(\mu^-,\mu^+)$ and the smuons $(\tilde{\mu}_\pm^-,\tilde{\mu}_\pm^+)$ which are assigned mass $M$. 
The $\grU(1)$ gauge superfield containing the photon $\gamma$ 
and the photino $\tilde{\gamma}$ is denoted by $\VV$. 
Employing the superspace conventions of Wess and 
Bagger \cite{Wess:1992cp}, the Lagrangian has the form
\be
\Lagr_{\mathrm{SQED}} \eq \Quarter \Bigbrk{\bigeval{WW}_{\theta^2} + \bigeval{\overline{W}\, \overline{W}}_{\bar{\theta}^2}} \nl
 +  \bigeval{ \Bigbrk{  \Phi_{e+}^\dagger e^{2e\VV} \Phi_{e+} + \Phi_{e-}^\dagger e^{-2e\VV} \Phi_{e-} } }_{\theta^2\bar{\theta}^2} 
 + m \Bigbrk{ \bigeval{ \Phi_{e+} \Phi_{e-} }_{\theta^2} + \bigeval{ \Phi_{e+}^\dagger \Phi^\dagger_{e-} }_{\bar{\theta}^2} } \nl
 +  \bigeval{ \Bigbrk{  \Phi_{m+}^\dagger e^{2e\VV} \Phi_{m+} + \Phi_{m-}^\dagger e^{-2e\VV} \Phi_{m-} } }_{\theta^2\bar{\theta}^2} 
 + M \Bigbrk{ \bigeval{ \Phi_{m+} \Phi_{m-} }_{\theta^2} + \bigeval{ \Phi_{m+}^\dagger \Phi^\dagger_{m-} }_{\bar{\theta}^2} }  \; ,
\ee
where the super fieldstrength is defined by $W_\alpha = - \quarter \bar{D}^2 D_\alpha \VV$ and $\overline{W}_\dalpha = - \quarter D^2 \bar{D}_\dalpha \VV$. After integrating out the auxiliary fields, the Lagrangian can be written as a kinetic term for the gauge fields
\be
 \Lagr_{\mathrm{gauge}} \eq
 - \Quarter F^{\mu\nu} F_{\mu\nu}
 + \iHalf \bar{\lambda} \gamma^\mu \partial_\mu \lambda
 \; ,
\ee
a part that contains the electronic particles
\be
 \Lagr_{\mathrm{electron}} \eq
   \bar{\psi}_e \bigbrk{ i \gamma^\mu \deriD_\mu + m } \psi_e
 + \phi_{e+}^\dagger \bigbrk{ \deriD^2 - m^2 }\phi_{e+}
 + \phi_{e-}^\dagger \bigbrk{ \deriD^2 - m^2 }\phi_{e-} \nl[1mm]
 + \sqrt{2}ie \bigbrk{ \phi_{e+} \bar{\psi}_e P_- \lambda - \phi_{e+}^\dagger \bar{\lambda} P_+ \psi_e
                     - \phi_{e-} \bar{\lambda} P_- \psi_e + \phi_{e-}^\dagger \bar{\psi}_e P_+ \lambda }
 \; ,
\ee
an analogous one for the muons, $\Lagr_{\mathrm{muon}}$, which is obtained by replacing the labels $e$ by $m$, and a part with contact interactions between the two families
\be
 \Lagr_{\mathrm{contact}} \eq
 - \frac{e^2}{2} \bigbrk{ \abs{\phi_{e+}}^2 - \abs{\phi_{e-}}^2 + \abs{\phi_{m+}}^2 - \abs{\phi_{m-}}^2 }^2
 \; .
\ee
For our notation and conventions, see \appref{app:conventions}.

\section{Scattering amplitudes}
\label{sec:As}

In order to find the spectrum of bound states of a particle of the electronic family and an anti-particle of the muonic family in \secref{sec:eigenstates}, we first compute the potential between any two of these particles. We deduce the potential energies from the non-relativistic limit of the tree-level scattering amplitudes which we compute from SQED Feynman rules. The amplitudes will allow us to calculate the bound state spectrum including all effects up to order $\alpha^4$ in the fine structure constant $\alpha = \frac{e^2}{4\pi}$.

In the next subsection, we will explicate the derivation of the potential from the amplitudes for the scattering of an electron and an anti-muon.  This scattering process is the only one that would exist for pure QED. The amplitudes and results for all other cases are listed in the two subsequent subsections.

\subsection{QED}

At tree-level the only diagram describing the scattering of an electron and an anti-muon involves the exchange of a photon:
\be
\raisebox{-7mm}{\includegraphics{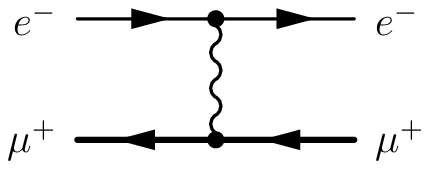}\hspace{4mm}}
= e^2 \bar{u}_e(q) \gamma^\mu u_e(p) \, \tilde \Delta_{\mu\nu}(p-q) \, \bar{v}_m(p') \gamma^\nu v_m(q') 
\ .
\ee
The in-going electron and anti-muon momenta are $p$ and $p'$ respectively.  The outgoing momenta are $q$ and $q'$.  
%
The spinors are 
\begin{equation}
u_e(p) = 
\left(
\begin{array}{r}
\sqrt{\sigma \cdot p} \, \xi_{ei} \\
\sqrt{\bar \sigma \cdot p} \, \xi_{ei}
\end{array}
\right)
\; , \; \; \;
u_e(q) = 
\left(
\begin{array}{r}
\sqrt{\sigma \cdot q} \, \xi_{eo} \\
\sqrt{\bar \sigma \cdot q} \, \xi_{eo}
\end{array}
\right) \ ,
\end{equation}
\begin{equation}
v_m(p') = 
\left(
\begin{array}{r}
\sqrt{\sigma \cdot p'} \, \eta_{mi} \\
-\sqrt{\bar \sigma \cdot p'} \, \eta_{mi}
\end{array}
\right) \; , \; \; \;
v_m(q') = 
\left(
\begin{array}{r}
\sqrt{\sigma \cdot q'} \, \eta_{mo} \\
-\sqrt{\bar \sigma \cdot q'} \, \eta_{mo}
\end{array}
\right) \ .
\end{equation}
We will work in the center of mass frame where 
\begin{equation}
p = (\sqrt{m^2 + {\vec p}^2}, {\vec p}) \; , \; \; \;
q = (\sqrt{m^2 + {\vec q}^2}, {\vec q}) \ , 
\end{equation}
\begin{equation}
p' = (\sqrt{M^2 + {\vec p}^2}, -{\vec p}) \; , \; \; \;
q' = (\sqrt{M^2 + {\vec q}^2}, -{\vec q}) \ .
\end{equation}

The first few terms in a non-relativistic expansion of the scattering amplitude are\footnote{
We are grateful to Tomas Rube and Jay Wacker for pointing out a mistake in this and similar formulas in a previous version of the manuscript. Before we worked in Feynman gauge where it would have been necessary to compute a one-loop diagram to fix an ambiguity in the non-relativistic potential \cite{Lindgren}.
}
\begin{eqnarray}
i {\mathcal M}
&=&
i \frac{4 M m e^2}{(\vec p - \vec q)^2} 
\xi_{eo}^\dagger  \eta_{mi}^\dagger
\biggsbrk{
 1 + \frac{1}{8} (\vec p + \vec q)^2
\biggbrk{ \frac{1}{m} + \frac{1}{M} }^2  
- \frac{(\vec p \, {}^2 - \vec q \, {}^2)^2}{(\vec p - \vec q)^2} \frac{1}{4Mm}
 \\
&& \hspace{8mm}
-\frac{i}{2}  (\vec p \times \vec q) \cdot \vec \sigma_e  \,
 \biggbrk{ \frac{1}{2m^2} + \frac{1}{Mm} }
+\frac{i}{2}  \, 
(\vec p \times \vec q) \cdot \vec \sigma_m 
\biggbrk{ \frac{1}{2M^2} + \frac{1}{Mm} }
\nonumber \\
&& \hspace{8mm}
+\frac{1}{4Mm} (\vec p - \vec q)^2 \vec \sigma_e \cdot \vec \sigma_m - \frac{1}{4Mm}
(\vec p - \vec q) \cdot \vec \sigma_e \, (\vec p - \vec q) \cdot \vec \sigma_m
 + \ldots
} \eta_{mo} \xi_{ei} \ . \nonumber
\end{eqnarray}
The ellipses denote terms that are higher order in the space-like momenta $\vec p$ and $\vec q$.
The plane wave states in quantum field theory are normalized to the Lorentz invariant quantity:
\be
\langle \vec p \, | \vec q \rangle = 2 \sqrt{m^2 + \vec q^{\, 2}} (2\pi)^3 \delta^{(3)} (\vec p- \vec q)
\ .
\ee
In non-relativistic quantum mechanics, in contrast, these plane wave states are typically normalized to $(2\pi)^3 \delta^{(3)} (\vec p - \vec q)$. To take into account the different normalizations, we define the non-relativistic scattering amplitude
\be
\mathcal{M}_{\rm NR} \equiv \frac{\mathcal M}{4 \bigsbrk{ (m^2 + \vec p^{\, 2}) (m^2 + \vec q^{\, 2}) (M^2 + \vec p^{\, 2})(M^2 + \vec q^{\, 2}) }^{1/4}} \ .
\ee
Taking into account the change in normalization, we find that
\be
 i \mathcal{M}_\mathrm{NR} \eq \frac{ie^2}{(\vec{p}-\vec{q})^2} \, \xi^\dagger_{eo} \xi^\dagger_{mo} \biggsbrk{
 1
 + \frac{1}{Mm} \left( \vec p \, {}^2 + \frac{(\vec p -\vec q)^2}{4} - \frac{(\vec p \cdot (\vec p - \vec q))^2}{(\vec p - \vec q)^2} \right)
 - \frac{1}{8} \biggbrk{\frac{1}{m}+\frac{1}{M}}^2 \, (\vec{p}-\vec{q})^2
 \nl \hspace{6mm}
 - \frac{i}{2} \biggbrk{\frac{1}{Mm}+\frac{1}{2m^2}} \, (\vec{p}\times\vec{q})\cdot\vec{\sigma}_e
 - \frac{i}{2} \biggbrk{\frac{1}{Mm}+\frac{1}{2M^2}} \, (\vec{p}\times\vec{q})\cdot\vec{\sigma}_m
 \nl \hspace{6mm}
 - \frac{1}{4Mm} \, (\vec{p}-\vec{q})^2 \, \vec{\sigma}_e \cdot\vec{\sigma}_m
 + \frac{1}{4Mm} \, (\vec{p}-\vec{q})\cdot\vec{\sigma}_e \, (\vec{p}-\vec{q})\cdot\vec{\sigma}_m
+ \ldots } \xi_{mi} \xi_{ei}  \ . 
\ee
We have changed the spinor $\eta$ of the anti-muon into a spinor $\xi$ as if it described a muon, $\eta = i\sigma^2 \xi^*$.

We would like to compare this scattering amplitude with the Born approximation result for a particle of position $\vec r$ and momentum $\vec p$ in non-relativistic quantum mechanics scattering off of a potential $V(\vec r, \vec p)$.  The Born approximation says that
\be
 \mathcal{M}_{\rm NR} = -\int d^3 \vec r \, e^{-i \vec q \cdot \vec r} V(\vec r, \vec p) e^{i \vec p \cdot  \vec r} \ ,
\ee 
for plane wave initial and final states.  We now Fourier transform the amplitude ${\mathcal M}_{\rm NR}$ with respect to $\vec p - \vec q$, keeping $\vec p$ as a variable. 
We find $\mbox{FT}({\mathcal M}_{\rm NR}) =  -V(\vec r, \vec p)$ and 
\begin{eqnarray}
\label{QEDpot}
V(\vec r, \vec p)&=& \frac{e^2}{4\pi} \biggsbrk{
-\frac{1}{r} 
%
- \frac{1}{Mm} \left( \frac{ \vec p \, {}^2}{2r} + \frac{ (\vec{r}\cdot\vec{p})^2}{2 r^3}
+ \pi \delta^{(3)}(\vec r) \right)
+ \frac{\pi}{2}  \delta^{(3)}  (\vec r) \biggbrk{ \frac{1}{m} + \frac{1}{M} }^2 
\nonumber \\
&& \hspace{8mm}
+ \frac{ \vec L \cdot \vec S_e}{r^3} \biggbrk{ \frac{1}{2m^2} + \frac{1}{M m} }
+ \frac{ \vec L \cdot \vec S_m}{r^3} \biggbrk{ \frac{1}{2M^2} + \frac{1}{M m} }
\nonumber \\
&& \hspace{8mm}
+\frac{1}{Mm} \biggbrk{
\frac{8 \pi}{3} \vec S_e \cdot \vec S_m \, \delta^{(3)}(\vec r) + \frac{3 \hat r \cdot \vec S_e \, \hat r \cdot \vec S_m-\vec S_e \cdot \vec S_m}{r^3}
} + \ldots
} \ .
\end{eqnarray}
All terms are understood to be normal ordered, i.e. when $\vec{p}$ and $\vec{L}$ are replaced by operators then they do \emph{not} act on the coordinate dependence of the potential. The result \eqref{QEDpot} is familiar up to subleading corrections in $1/M$. The first term is the Coulomb attraction. The second term is the orbit-orbit, also referred to as the current-current, coupling. The third term is the Darwin term.  The fourth term is the spin-orbit coupling of the electron.  The fifth term is the spin orbit coupling of the muon.  The last term is the hyperfine coupling between the spin of the electron and the spin of the muon.  For a hydrogenic orbital, the expectation values of $\vev{1/ r}$ and $\vev{\vec p}$ scale as $\alpha \mu$.  Thus, this non-relativistic expansion of the effective potential is also an expansion in the fine structure constant.  The Coulomb interaction is of order $\alpha^2 \mu$ and the other terms are suppressed by an additional power of $\alpha^2$.

\subsection{Bosonic amplitudes}

In SQED, an electron anti-muon bound state mixes with a selectron anti-smuon bound state through photino exchange.  To calculate the energy spectrum, there are a number of additional scattering diagrams that must be computed.

\begin{itemize}
	\item {\mathversion{bold} $e^- \mu^+ \rightarrow \tilde{e}^-_\pm \tilde{\mu}^+_\pm$}
\begin{flalign}
& \label{MNRemusesmu}
\raisebox{-7mm}[14mm][10mm]{\includegraphics{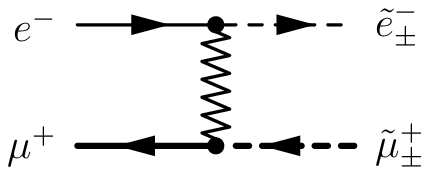}\hspace{4mm}}
= -2ie^2 \, \frac{\bar{v}_m(p') (\slashed{p}-\slashed{q}) P_\pm u_e(p)}{(p-q)^2} && \\
& \nn
 i \mathcal{M}_\mathrm{NR} = \frac{ie^2}{(\vec{p}-\vec{q})^2} \, \eta^\dagger_{mi} \biggsbrk{
 - \frac{(\vec{p}-\vec{q})\cdot\vec{\sigma}}{2\sqrt{Mm}}
 \mp \frac{1}{8} \frac{M-m}{(Mm)^{3/2}} \, (\vec{p}-\vec{q})^2
 \mp \frac{i}{4} \frac{M+m}{(Mm)^{3/2}} \, (\vec{p}\times\vec{q})\cdot\vec{\sigma}
 } \xi_{ei} \\[1mm]
& \nn
  V = - \frac{e^2}{4\pi} \xi^\trans_{mi} \,i\sigma^2 \biggsbrk{
     - \frac{i}{2\sqrt{Mm}} \frac{\vec{r}\cdot\vec{\sigma}}{r^3}
     \pm \frac{\pi}{2} \frac{M-m}{(Mm)^{3/2}} \, \delta^{(3)}(\vec{r})
     \pm \frac{1}{4} \frac{M+m}{(Mm)^{3/2}} \frac{\vec{L}\cdot\vec{\sigma}}{r^3}
  } \xi_{ei} \\[8mm] \nn
\end{flalign}

	\item {\mathversion{bold} $\tilde{e}^-_\pm \tilde{\mu}^+_\pm \rightarrow e^- \mu^+$}
\begin{flalign}
& \label{MNRsesmuemu}
\raisebox{-7mm}[14mm][10mm]{\includegraphics{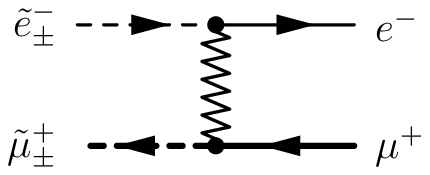}\hspace{4mm}}
= 2ie^2 \, \frac{\bar{u}_e(q) (\slashed{p}-\slashed{q}) P_\pm v_m(q')}{(p-q)^2} && \\
& \nn
 i \mathcal{M}_\mathrm{NR} = \frac{ie^2}{(\vec{p}-\vec{q})^2} \, \xi^\dagger_{eo} \biggsbrk{
   \frac{(\vec{p}-\vec{q})\cdot\vec{\sigma}}{2\sqrt{Mm}}
 \mp \frac{1}{8} \frac{M-m}{(Mm)^{3/2}} \, (\vec{p}-\vec{q})^2
 \mp \frac{i}{4} \frac{M+m}{(Mm)^{3/2}} \, (\vec{p}\times\vec{q})\cdot\vec{\sigma}
 } \eta_{mo} \\[1mm]
& \nn
  V = - \frac{e^2}{4\pi} \xi^\dagger_{eo} \biggsbrk{
     - \frac{i}{2\sqrt{Mm}} \frac{\vec{r}\cdot\vec{\sigma}}{r^3}
     \mp \frac{\pi}{2} \frac{M-m}{(Mm)^{3/2}} \, \delta^{(3)}(\vec{r})
     \mp \frac{1}{4} \frac{M+m}{(Mm)^{3/2}} \frac{\vec{L}\cdot\vec{\sigma}}{r^3}
  } i\sigma^2\, \xi^*_{mo} \\[8mm] \nn
\end{flalign}

	\item {\mathversion{bold} $\tilde{e}^-_\pm \tilde{\mu}^+_\pm \rightarrow \tilde{e}^-_\pm \tilde{\mu}^+_\pm$}
\begin{flalign}
&
\raisebox{-7mm}[14mm][10mm]{\includegraphics{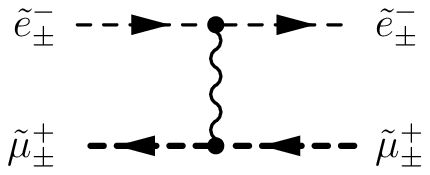}\hspace{4mm}}
= e^2 (p+q)^\mu \tilde \Delta_{\mu \nu}(p-q)(p'+q')^\nu && \\
&
\raisebox{-7mm}[10mm][10mm]{\includegraphics{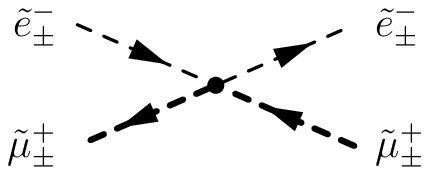}\hspace{4mm}}
= -ie^2 \\
& \nn
 i \mathcal{M}_\mathrm{NR} = \frac{ie^2}{(\vec{p}-\vec{q})^2} \biggsbrk{
 1
 + \frac{1}{Mm} \left( \vec p \, {}^2 + \frac{(\vec p -\vec q)^2}{4} - \frac{(\vec p \cdot (\vec p - \vec q))^2}{(\vec p - \vec q)^2} \right)
%
 - \frac{1}{2Mm} \, (\vec{p}-\vec{q})^2
 } \\[1mm]
& \nn
  V = - \frac{e^2}{4\pi} \biggsbrk{
     \frac{1}{r}
     + \frac{1}{Mm} \left( \frac{ \vec p \, {}^2}{2r} + \frac{ (\vec{r}\cdot\vec{p})^2}{2 r^3}
+ \pi \delta^{(3)}(\vec r) \right)
%
     - \frac{2\pi}{Mm} \delta^{(3)}(\vec{r})
  } \\[8mm] \nn
\end{flalign}

	\item {\mathversion{bold} $\tilde{e}^-_\pm \tilde{\mu}^+_\mp \rightarrow \tilde{e}^-_\pm \tilde{\mu}^+_\mp$}
\begin{flalign}
&
\raisebox{-7mm}[14mm][10mm]{\includegraphics{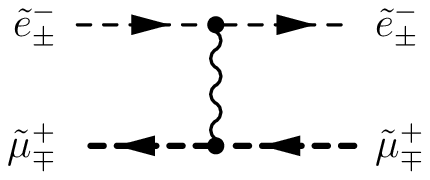}\hspace{4mm}}
= e^2 (p+q)^\mu \tilde \Delta_{\mu\nu}(p-q) (p'+q')^\nu && \\
&
\raisebox{-7mm}[10mm][10mm]{\includegraphics{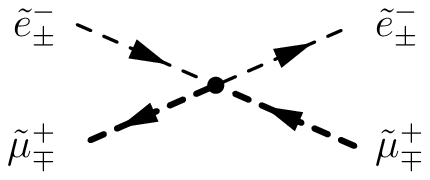}\hspace{4mm}}
= ie^2 \\
& \nn
 i \mathcal{M}_\mathrm{NR} = \frac{ie^2}{(\vec{p}-\vec{q})^2} \biggsbrk{
 1
 + \frac{1}{Mm} \left( \vec p \, {}^2 + \frac{(\vec p -\vec q)^2}{4} - \frac{(\vec p \cdot (\vec p - \vec q))^2}{(\vec p - \vec q)^2} \right)
%
 } \\[1mm]
& \nn
  V = - \frac{e^2}{4\pi} \biggsbrk{
     \frac{1}{r}
      + \frac{1}{Mm} \left( \frac{ \vec p \, {}^2}{2r} + \frac{ (\vec{r}\cdot\vec{p})^2}{2 r^3}
+ \pi \delta^{(3)}(\vec r) \right)
     %
  }
\end{flalign}

\end{itemize}

\newpage
\subsection{Fermionic amplitudes}

In addition to bosonic bound states in SQED, there are fermionic bound states involving an electron and anti-smuon or selectron and anti-muon.

\begin{itemize}
	\item {\mathversion{bold} $e^- \tilde{\mu}^+_\pm \rightarrow e^- \tilde{\mu}^+_\pm$}
\begin{flalign}
&
\raisebox{-7mm}[14mm][10mm]{\includegraphics{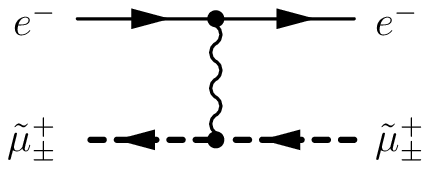}\hspace{4mm}}
= e^2 \bar{u}_e(q) \gamma^\mu \tilde \Delta_{\mu\nu}(p-q) (p'+q')^\nu
u_e(p)  && \\
& \nn
 i \mathcal{M}_\mathrm{NR} = \frac{ie^2}{(\vec{p}-\vec{q})^2} \, \xi^\dagger_{eo} \biggsbrk{
 1
+ \frac{1}{Mm} \left( \vec p \, {}^2 + \frac{(\vec p -\vec q)^2}{4} - \frac{(\vec p \cdot (\vec p - \vec q))^2}{(\vec p - \vec q)^2} \right) \\
%
& \nn \hspace{30mm} - \frac{1}{4} \biggbrk{\frac{1}{Mm}+\frac{1}{2m^2}} \, (\vec{p}-\vec{q})^2 
 - \frac{i}{2} \biggbrk{\frac{1}{Mm}+\frac{1}{2m^2}} \, (\vec{p}\times\vec{q})\cdot\vec{\sigma}
 } \, \xi_{ei} \\[1mm]
& \nn
  V = - \frac{e^2}{4\pi} \xi^\dagger_{eo} \biggsbrk{
     \frac{1}{r}
     + \frac{1}{Mm} \left( \frac{ \vec p \, {}^2}{2r} + \frac{ (\vec{r}\cdot\vec{p})^2}{2 r^3}
+ \pi \delta^{(3)}(\vec r) \right)
\\
   %
  &\nn \hspace{50mm}   - \pi \biggbrk{\frac{1}{Mm}+\frac{1}{2m^2}} \, \delta^{(3)}(\vec{r})
     - \frac{1}{2} \biggbrk{\frac{1}{Mm}+\frac{1}{2m^2}} \frac{\vec{L}\cdot\vec{\sigma}}{r^3}
  } \xi_{ei} \\[8mm] \nn
\end{flalign}

\item {\mathversion{bold} $\tilde{e}^-_\pm \mu^+ \rightarrow \tilde{e}^-_\pm \mu^+$}
\begin{flalign}
&
\raisebox{-7mm}[14mm][10mm]{\includegraphics{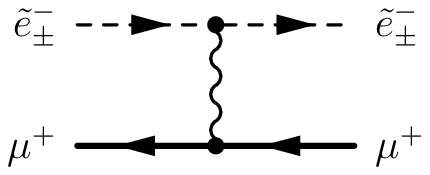}\hspace{4mm}}
= e^2 \bar{v}_m(p') \gamma^\mu \tilde \Delta_{\mu\nu}(p-q) (p+q)^\nu
v_m(q') && \\
& \nn
 i \mathcal{M}_\mathrm{NR} = \frac{ie^2}{(\vec{p}-\vec{q})^2} \, \eta^\dagger_{mi} \biggsbrk{
 1
 + \frac{1}{Mm} \left( \vec p \, {}^2 + \frac{(\vec p -\vec q)^2}{4} - \frac{(\vec p \cdot (\vec p - \vec q))^2}{(\vec p - \vec q)^2} \right) \\
 %
& \nn
 \hspace{30mm}
 - \frac{1}{4} \biggbrk{\frac{1}{Mm}+\frac{1}{2M^2}} \, (\vec{p}-\vec{q})^2 
 + \frac{i}{2} \biggbrk{\frac{1}{Mm}+\frac{1}{2M^2}} \, (\vec{p}\times\vec{q})\cdot\vec{\sigma}
 } \, \eta_{mo} \\[1mm]
& \nn
  V = - \frac{e^2}{4\pi} \xi^\dagger_{mo} \biggsbrk{
     \frac{1}{r}
       + \frac{1}{Mm} \left( \frac{ \vec p \, {}^2}{2r} + \frac{ (\vec{r}\cdot\vec{p})^2}{2 r^3}
+ \pi \delta^{(3)}(\vec r) \right) 
\\
& \nn \hspace{50mm}
     %
     - \pi \biggbrk{\frac{1}{Mm}+\frac{1}{2M^2}} \, \delta^{(3)}(\vec{r})
     - \frac{1}{2} \biggbrk{\frac{1}{Mm}+\frac{1}{2M^2}} \frac{\vec{L}\cdot\vec{\sigma}}{r^3}
  } \xi_{mi} \\[8mm] \nn
\end{flalign}

\item {\mathversion{bold} $e^- \tilde{\mu}^+_\pm \rightarrow \tilde{e}^-_\mp \mu^+$}
\begin{flalign}
& \label{MNResmusemu}
\raisebox{-7mm}[14mm][10mm]{\includegraphics{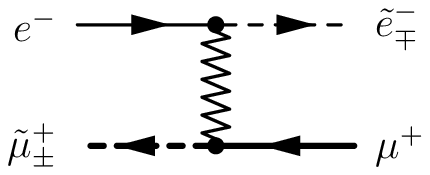}\hspace{4mm}}
= 2ie^2 \, \frac{u^\trans_e(p) C (\slashed{p}-\slashed{q}) P_\pm v_m(q') }{(p-q)^2} && \\
& \nn
 i \mathcal{M}_\mathrm{NR} = \frac{ie^2}{(\vec{p}-\vec{q})^2} \, \xi^\trans_{ei} \, i\sigma^2 \biggsbrk{
 \pm \frac{(\vec{p}-\vec{q})\cdot\vec{\sigma}}{2\sqrt{Mm}}
 + \frac{1}{8} \frac{M+m}{(Mm)^{3/2}} \, (\vec{p}-\vec{q})^2
 - \frac{i}{4} \frac{M+m}{(Mm)^{3/2}} \, (\vec{p}\times\vec{q})\cdot\vec{\sigma}
 } \, \eta_{mo} \\[1mm]
& \nn
  V = - \frac{e^2}{4\pi} \xi^\dagger_{mo} \biggsbrk{
     \mp \frac{i}{2\sqrt{Mm}} \frac{\vec{r}\cdot\vec{\sigma}}{r^3}
     - \frac{\pi}{2} \frac{M+m}{(Mm)^{3/2}} \, \delta^{(3)}(\vec{r})
     - \frac{1}{4} \frac{M+m}{(Mm)^{3/2}} \frac{\vec{L}\cdot\vec{\sigma}}{r^3}
  } \xi_{ei} \\[8mm] \nn
\end{flalign}

\item {\mathversion{bold} $\tilde{e}^-_\pm \mu^+ \rightarrow e^- \tilde{\mu}^+_\mp$}
\begin{flalign}
& \label{MNRsemuesmu}
\raisebox{-7mm}[14mm][10mm]{\includegraphics{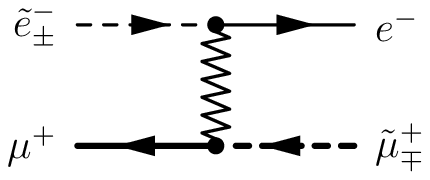}\hspace{4mm}}
= -2ie^2 \, \frac{\bar{u}_e(q) (\slashed{p}-\slashed{q}) C P_\pm \bar{v}^\trans_m(p') }{(p-q)^2} && \\
& \nn
 i \mathcal{M}_\mathrm{NR} = \frac{ie^2}{(\vec{p}-\vec{q})^2} \, \xi^\dagger_{eo} \biggsbrk{
 \mp \frac{(\vec{p}-\vec{q})\cdot\vec{\sigma}}{2\sqrt{Mm}}
 + \frac{1}{8} \frac{M+m}{(Mm)^{3/2}} \, (\vec{p}-\vec{q})^2
 + \frac{i}{4} \frac{M+m}{(Mm)^{3/2}} \, (\vec{p}\times\vec{q})\cdot\vec{\sigma}
 } i\sigma^2\, \eta^*_{mi} \\[1mm]
& \nn
  V = - \frac{e^2}{4\pi} \xi^\dagger_{eo} \biggsbrk{
     \mp \frac{i}{2\sqrt{Mm}} \frac{\vec{r}\cdot\vec{\sigma}}{r^3}
     - \frac{\pi}{2} \frac{M+m}{(Mm)^{3/2}} \, \delta^{(3)}(\vec{r})
     - \frac{1}{4} \frac{M+m}{(Mm)^{3/2}} \frac{\vec{L}\cdot\vec{\sigma}}{r^3}
  } \xi_{mi}
\end{flalign}

\end{itemize}

\section{Mixing matrices}
\label{sec:mixing}

In the previous section we have derived the non-relativistic expansions of the potential energy $V(\vec{r},\vec{p})$ between an electron and an anti-muon or their super partners in terms of the relative coordinate $\vec{r}$ and the relative momentum $\vec{p}$. To find the effective quantum mechanical description of this system, we also need to expand the kinetic energy of these particles
\be
  E_{\mathrm{kin}} = \sqrt{m^2 + p^2_e} + \sqrt{M^2 + p^2_m} - (m + M) 
\ee
to the same order, i.e. to fourth order in the momenta. Then the Hamiltonian becomes
\be \label{eqn:Hamiltonian}
  H = \frac{\vec{p}\,{}^2}{2\mu} - \frac{1}{8}\Bigbrk{\frac{1}{m^3} + \frac{1}{M^3}}\vec{p}\,{}^4 + V(\vec{r},\vec{p})
    = \frac{\vec{p}\,{}^2}{2\mu} - \frac{\alpha}{r} + H_{\mathrm{int}}(\vec{r},\vec{p}) \; ,
\ee
where $V$ and $H_{\mathrm{int}}$ are matrices acting on the various ``spin'' states $\ket{s_e s_m}$ where $s\in\{\uparrow,\downarrow,+,-\}$.  The components of $V$ are the several potentials given in \secref{sec:As}. In \eqref{eqn:Hamiltonian} we have singled out the non-relativistic kinetic energy and the Coulomb potential, and denote all other terms by $H_{\mathrm{int}}$.

We are interested in the bound state spectrum of this system. Without the interactions $H_{\mathrm{int}}$, the solutions would be the familiar hydrogenic bound states $\ket{n l m_l}$ with the Bohr energies $E_n = -\mu \alpha^2 / 2 n^2$, see \appref{sec:Coulomb-problem}. Our task now is to determine the $\alpha^4$ corrections to this spectrum, which have two different sources.  The first one is first order degenerate perturbation theory. Most of the terms in the scattering amplitudes are of order $\alpha^4$ and lead to mixing between the states in the highly degenerate levels of a given $n$ and $l$.  

There are a handful of terms in the scattering amplitudes that are of order $\alpha^3$, namely the first terms in (\ref{MNRemusesmu}), (\ref{MNRsesmuemu}), (\ref{MNResmusemu}), and (\ref{MNRsemuesmu}).  Naively, these terms should dominate the $\alpha^4$ contributions.  However, it turns out that
\be
\Big\langle n,l,m_l \Big| \frac{\vec r}{r^3} \Big| n,l',m_l' \Big\rangle = 0 \; .
\label{intervanish}
\ee
Thus, these terms do not contribute at the level of first order perturbation theory.  However, as was noted in \cite{Buchmuller:1981bp}, they can and do contribute at second order. Recall the formula for the energy corrections
\be
E^{(2)}_i = \sum_{j \neq i} \frac{|\bra{i} H_{\mathrm{int}} \ket{j} |^2}{E_i^{(0)} - E_j^{(0)}} \; ,
\label{secordpert}
\ee
where $E_i^{(0)}$ are the eigen-energies of the bare Hamiltonian.  For $\bra{i} H_{\mathrm{int}} \ket{j}$ of order $\alpha^3$ and $E_i^{(0)}$ of order $\alpha^2$ the second order corrections will be of order $\alpha^4$. The sum in \eqref{secordpert} should be carried over both discrete and continuum states of the hydrogen atom. To carry out the sum, we will make use of Schwinger's Coulomb Green's function \cite{Schwinger:1964}.

At first glance, the diagonalization problem of the $\ket{nlm_l}$ states seems formidable. For a given $n$, we have $n$ different $l$'s, for each $l$, we have $2l+1$ different $m_l$'s, and for each $m_l$, we have 16 different ``spins'' $\ket{s_e s_m}$ all of the same energy. As it turns out, states of different $l$ do not mix. Moreover, the total angular momentum in the $z$ direction is a good quantum number. The largest matrix we will need to diagonalize is $6 \times 6$.  

That the total $z$-component of angular momentum is conserved is obvious, but that states of different $l$ do not mix is surprising.  Both the second order perturbative corrections and the hyperfine interaction have the potential to mix an $l$ state with an $l+2$ state.  For the hyperfine interaction, one can check explicitly that the overlap integral
\be
\Big\langle n, l+2, m_l \Big| \frac{ r_i r_j}{r^3} \Big| n, l, m_l' \Big\rangle = 0 
\ee
vanishes. Another integral, which we discuss in Appendix \ref{app:secorderpert}, guarantees that there is no mixing of states with different $l$ in second order perturbation theory.

In the appendices, we discuss separately the contributions from first order degenerate perturbation theory to the $l=0$ and $l>0$ cases.  The reason for the separation is that the Dirac delta functions in the scattering amplitudes are only important for $l=0$ states while the spin-orbit interactions only contribute when $l>0$. In the appendices, we also will calculate the contribution from second order perturbation theory. Below, we present the final result for the mixing matrices.

\subsection{Overall shift}
\label{sec:energyshift}

The relativistic correction to the kinetic energy as well as a term $\sim\frac{1}{Mm}$ in the potential $V$ do not depend on the spins of the particles. Therefore, these terms lead to an overall shift of the levels specified by $n$ and $l$. We can compute this shift independently from the splitting. It is given by the expectation values of the following terms in the $\ket{nlm_l}$ basis:
\be
  \label{eqn:energyshift}
  \delta E_{nl} \eq - \frac{\alpha}{Mm} \biggvev{
  \frac{\vec{p}\,{}^2}{2r} + \frac{(\vec{r}\cdot\vec{p})^2}{2r^3} + \pi \delta^{(3)}(\vec r) 
  } - \frac{1}{8} \biggbrk{\frac{1}{m^3}+\frac{1}{M^3}} \vev{\vec{p}\,{}^4} \nln
                \eq - \frac{\mu\alpha^4}{n^4} \biggsbrk{ \frac{n}{2l+1} - \frac{3}{8}  + \frac{\mu^2}{8Mm} } \; .
\ee

\subsection{$l=0$-states}

There is an eight dimensional space of bosonic bound states with $l=0$:
\be
  && \ket{\mathrm{in}} \in \Big\{
   \ket{\uparrow \uparrow},\,
   \ket{\downarrow \uparrow},\,
   \ket{\uparrow \downarrow},\,
   \ket{\downarrow \downarrow},\, 
   \ket{++},\,
   \ket{--},\,
   \ket{+-},\,
   \ket{-+}
   \Big\} \; .
\ee
(Because the $l=0$ sector is already relatively small, we do not take advantage of the fact that angular momentum in the $z$-direction is a good quantum number.)
The first entry of the state describes the electronic portion of the bound state, whether the electron is spin up or down, or whether the selectron comes from the $+$ or $-$ chiral superfield. The second entry describes the muonic portion. The Hamiltonian to order $\alpha^4$ for these states takes the form $H = E_{n} + \delta E_{n0} + M_b$. Assembling the contributions from both first and second order perturbation theory, the mixing matrix for these states is
\be \label{eqn:mixing-b-l=0}
  M_b = \frac{\mu \alpha^4}{2n^3} \matr{ccc}{
  A & B^\trans & 0 \\
  B & C & 0 \\
  0 & 0 & 0
  } 
\ee
with
\be
A = \matr{cccc}{
  1 & 0 & 0 & 0 \\[2mm]
  0 & \frac{M^2+m^2}{(M+m)^2} & \frac{2Mm}{(M+m)^2} & 0 \\[2mm]
  0 & \frac{2Mm}{(M+m)^2} & \frac{M^2+m^2}{(M+m)^2} & 0 \\[2mm]
  0 & 0 & 0 & 1
 } \; ,
\ee
\be
B = \frac{\sqrt{Mm}(M-m)}{(M+m)^2} \matr{cccc}{
  0 & -1 & 1 & 0 \\
  0 & 1 & -1 & 0
 } \; ,
\ee
\be
C = \frac{2Mm}{(M+m)^2} \matr{cccc}{
  1 & -1 \\
  -1 & 1
 } \; .
\ee
A curious observation is that $M_b^2 = \frac{\mu \alpha^4}{2n^3} M_b$. Note that the $\ket{+-}$ and $\ket{-+}$ states decouple from the other six states; it remains to diagonalize a $6 \times 6$ matrix.

By supersymmetry, there is also an eight dimensional space of fermionic bound states with $l=0$:
\be
  && \ket{\mathrm{in}} \in \Big\{
   \ket{\uparrow +},\,
   \ket{- \uparrow},\,
   \ket{\downarrow +},\,
   \ket{- \downarrow},\,
   \ket{\uparrow -},\,
   \ket{+ \uparrow},\,
   \ket{\downarrow -},\,
   \ket{+ \downarrow}
   \Big\} \; ,
\ee
where the Hamiltonian takes the form $H = E_{n} + \delta E_{n0} + M_f + \order(\alpha^5)$. The mixing matrix in this case breaks up into a bunch of $2 \times 2$ blocks:
\be \label{eqn:mixing-f-l=0}
  M_f = \frac{\mu \alpha^4}{2n^3} \matr{cc}{
  D & 0 \\
  0 & D
  }
\ee
with
\be
D = \frac{1}{M+m} \matr{cccc}{
  M & \sqrt{M m} & 0 & 0 \\
  \sqrt{M m} & m & 0 & 0 \\
  0 & 0 & M & \sqrt{M m} \\
  0 & 0 & \sqrt{M m} & m
 } \; .
\ee
Note, for example, that the $\ket{\uparrow +}$ state mixes only with the $\ket{- \uparrow}$ state. Again we have $M_f^2 = \frac{\mu \alpha^4}{2n^3} M_f$.

\subsection{$l>0$-states}
\label{sec:mixing-l>0}

As explained above, there is no mixing between states with different $l$. Therefore, we fix the orbital angular momentum to some $l>0$. Furthermore, it is convenient to split this space, which contains $8\times (2l+1)$ bosonic states and $8 \times (2l+1)$ fermionic states, into closed subspaces of states with given $z$-component, $m_j$, of the total angular momentum. The bosonic sector of such a subspace is spanned by the states:
\be
  && \ket{\mathrm{in}} \in \Big\{
   \ket{l\,m_l-1, \uparrow \uparrow},\,
   \ket{l\,m_l, \downarrow \uparrow},\,
   \ket{l\,m_l, \uparrow \downarrow},\,
   \ket{l\,m_l+1, \downarrow \downarrow},\, 
  \nln && \hspace{15mm}
   \ket{l\,m_l, ++},\,
   \ket{l\,m_l, --},\,
   \ket{l\,m_l, +-},\,
   \ket{l\,m_l, -+}
   \Big\} \; .
   \label{bosonicin}
\ee
There are $2l+3$ such subspaces labeled by $m_j = m_l=-l-1,-l,\ldots,l+1$ where $j = l-1, l$, or $l$. States in the set \eqref{bosonicin} with magnetic quantum number outside the range $-l,...,l$ are understood to be absent. Thus the dimensions of these subspaces are $1,7,8,8,\ldots,8,7,1$.

The Hamiltonian acting on these states can be written to order $\alpha^4$ as $H = E_{n} + \delta E_{nl} + M_b$. The mixing matrix takes the form
\be \label{eqn:mixing-b-l>0}
  M_b = \frac{\mu \alpha^4}{2l(l+1)(2l+1) n^3} \matr{ccc}{
  A & B^\trans & 0 \\
  B & C & 0 \\
  0 & 0 & 0
  } \; .
\ee
In the leptonic sector it is given by
\be
A = \matr{cccc}{
  m_l-1                    & \frac{M}{M+m} c_{l,-m_l} & \frac{m}{M+m} c_{l,-m_l} & 0 \\[2mm]
  \frac{M}{M+m} c_{l,-m_l} & -\frac{M-m}{M+m} m_l     & 0                        & \frac{m}{M+m} c_{lm_l} \\[2mm]
  \frac{m}{M+m} c_{l,-m_l} & 0                        & \frac{M-m}{M+m} m_l      & \frac{M}{M+m} c_{lm_l} \\[2mm]
  0                        & \frac{m}{M+m} c_{lm_l}   & \frac{M}{M+m} c_{lm_l}   & -m_l-1
 }
\ee
and the mixing between leptons and sleptons is given by
\be
B = \frac{\sqrt{m M}}{M+m} \matr{cccc}{
  -c_{l,-m_l} &  m_l &  m_l &  c_{lm_l} \\[2mm]
   c_{l,-m_l} & -m_l & -m_l & -c_{lm_l}
 }
\ee
where $c_{lm_l} = \sqrt{(l-m_l)(l+m_l+1)}$. There is no interaction among the sleptons, $C=0$.

In the fermionic sector, the states have half-integer total magnetic quantum number $m_j = m_l +\half$ where the range of $m_l$ is $-l-1,-l,\ldots,l$. The corresponding $2l+2$ subspaces for $j=l\pm 1/2$ have dimensions $4,8,8,\ldots,8,4$ and are spanned by
\be
  && \ket{\mathrm{in}} \in \Big\{
   \ket{l\,m_l, \uparrow +},\,
   \ket{l\,m_l, - \uparrow},\,
   \ket{l\,m_l+1, \downarrow +},\,
   \ket{l\,m_l+1, - \downarrow},\,
  \nln && \hspace{15mm}
   \ket{l\,m_l, \uparrow -},\,
   \ket{l\,m_l, + \uparrow},\,
   \ket{l\,m_l+1, \downarrow -},\,
   \ket{l\,m_l+1, + \downarrow}
   \Big\} \; .
   \label{fermionicin}
\ee
For the mixing matrix in this sector, we find
\be \label{eqn:mixing-f-l>0}
  M_f = \frac{\mu \alpha^4}{2l(l+1)(2l+1) n^3} \matr{cc}{
  D & 0 \\
  0 & D
  }
\ee
with
\be
D = \matr{cccc}{
  \frac{M}{M+m}\, m_l               & \frac{\sqrt{M m}}{M+m}\, m_l      & \frac{M}{M+m}\, c_{lm_l}          & \frac{\sqrt{M m}}{M+m}\, c_{lm_l} \\[2mm]
  \frac{\sqrt{M m}}{M+m}\, m_l      & \frac{m}{M+m}\, m_l               & \frac{\sqrt{M m}}{M+m}\, c_{lm_l} & \frac{m}{M+m}\, c_{lm_l} \\[2mm]
  \frac{M}{M+m}\, c_{lm_l}          & \frac{\sqrt{M m}}{M+m}\, c_{lm_l} & -\frac{M}{M+m} (m_l+1)            & -\frac{\sqrt{M m}}{M+m} (m_l+1) \\[2mm]
  \frac{\sqrt{M m}}{M+m}\, c_{lm_l} & \frac{m}{M+m}\, c_{lm_l}          & -\frac{\sqrt{M m}}{M+m} (m_l+1)   & -\frac{m}{M+m} (m_l+1)
 } \; .
\ee

\section{Energy splittings and eigenstates}
\label{sec:eigenstates}

At order $\alpha^2$ in the fine structure constant, the energy spectrum is given by the $16n^2$-fold degenerate Bohr levels
\be
  E_{nlm_ls_es_m} = E_{n} = -\frac{\mu \alpha^2}{2n^2} \; .
\ee
They receive a spin independent shift $\delta E_{nl}$ at order $\alpha^4$, which we have already computed in \secref{sec:energyshift}. In this section we calculate the additional splittings of these levels and find the energy eigenstates. The splitting energies and the eigenstates are given by the eigenvalues and eigenvectors of the mixing matrices $M_b$ and $M_f$ computed above. Because spherically symmetric states ($l=0$) and asymmetric ones ($l>0$) split up differently into two and three levels, respectively (see \figref{fig:spectrum} on page \pageref{fig:spectrum}), we discuss these two cases separately.

\begin{figure}
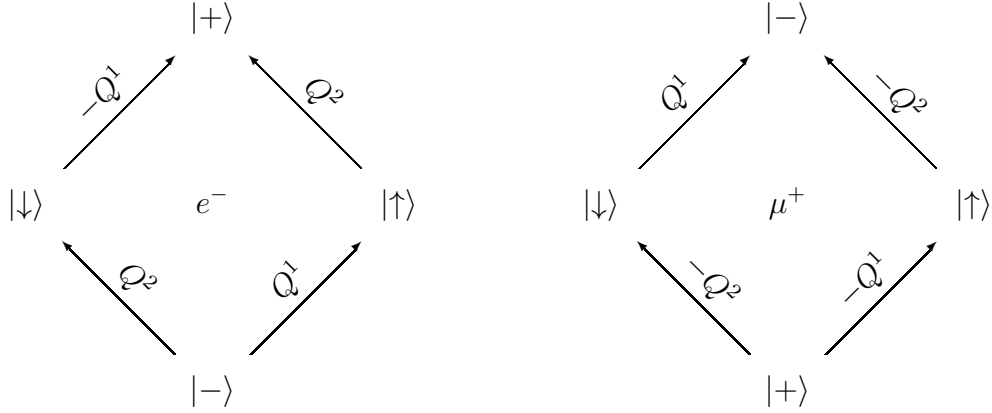
%
\begin{center}
\begin{tabular}{cp{16mm}c}
\begin{diagram}
                 &              & \ket{+} &             & \\
                 & \ruTo^{-Q^1} &         & \luTo^{Q^2} & \\
\ket{\downarrow} &              & e^-     &             & \ket{\uparrow} \\
                 & \luTo^{Q^2}  &         & \ruTo^{Q^1} & \\
                 &              & \ket{-} &             &
\end{diagram}
& &
\begin{diagram}
                 &              & \ket{-} &              & \\
                 & \ruTo^{Q^1}  &         & \luTo^{-Q^2} & \\
\ket{\downarrow} &              & \mu^+   &              & \ket{\uparrow} \\
                 & \luTo^{-Q^2} &         & \ruTo^{-Q^1} & \\
                 &              & \ket{+} &              &
\end{diagram}
\end{tabular}
\end{center}
\caption{\textbf{Action of supercharges.} The action of the $Q^\alpha$ is indicated by the arrows. Additionally there is factor of $\sqrt{m}$ or $\sqrt{M}$ when acting on electrons or muons, respectively. The action of $Q_\alpha^\dagger$ is the inverse of the action of $Q^\alpha$.}
\label{fig:action-of-supercharges}
\end{figure}

We organize the eigenstates that remain degenerate at order $\alpha^4$ into multiplets of the underlying supersymmetry algebra
\be
  &&
  \comm{J_a}{J_b} = i \levi_{abc} J_c
  \comma
  \comm{J_a}{Q^\alpha} = \half (\sigma_a)^\alpha{}_\beta Q^\beta
  \comma
  \comm{J_a}{Q^\dagger_\alpha} = - \half (\sigma_a)_\alpha{}^\beta Q^\dagger_\beta
  \; , \\[2mm]
  &&
  \comm{J_a}{H} = 0
  \comma
  \comm{Q^\alpha}{H} = 0
  \comma  
  \acomm{Q^\alpha}{Q^\dagger_\beta} = H \delta^\alpha_\beta 
  \; ,
\ee
where $\vec{J} = \vec{L} + \vec{S}_e + \vec{S}_m$ is the total angular momentum operator, $Q^\alpha$, $\alpha=1,2$, are the supercharges, and $H$ is the Hamiltonian. The action of the supercharges on states to zeroth order in $\alpha$ is depicted in \figref{fig:action-of-supercharges}. To this order they anti-commute to the rest energy $m+M$.
We denote super multiplets by $\mathcal{R}_j$ where $j=0,\half,1,\ldots$ refers to the total $\grSU(2)$ spin of the highest 
submultiplet, i.e. the one whose states are annihilated by the supercharges $Q_\alpha^\dagger$. In terms of spin-$j$ multiplets $V_j$ of $\grSU(2)$, the super multiplet $\mathcal{R}_j$ is built from $V_{j-1/2}\oplus 2 V_{j} \oplus V_{j+1/2}$ for $j>1/2$ and from $2 V_{0} \oplus V_{1/2}$ for $j=0$. The dimension of $\mathcal{R}_j$ is $(8j+4)$.

The energy eigenstates depend on the mass ratio $\tau \equiv \frac{m}{M}$.

\subsection{$l=0$-states}

The mixing matrices that need to be diagonalized in this case are given in \eqref{eqn:mixing-b-l=0} and \eqref{eqn:mixing-f-l=0} for the bosonic and fermionic bound states, respectively. We find that there are 4 bosonic and 4 fermionic states with eigenvalue $\Delta E = 0$, and 4 bosonic and 4 fermionic states with eigenvalue $\Delta E=\frac{\mu\alpha^4}{2n^3}$. It turn out that the former states constitute two super multiplets $\mathcal{R}_0$, while the latter ones fill one $\mathcal{R}_{1/2}$. See the $l=0$ column of \figref{fig:spectrum} where these results are visualized.

The energy eigenstates in the first $\mathcal{R}_0$ with $\Delta E = 0$ are given by
\be
  V_0:      && \ket{-+} \\
  V_{1/2}:  && \sqrt{\tfrac{\tau\vphantom{1}}{1+\tau}} \, \ket{\uparrow +}   - \sqrt{\tfrac{1}{1+\tau}} \, \ket{-\uparrow}
        \comma \sqrt{\tfrac{\tau\vphantom{1}}{1+\tau}} \, \ket{\downarrow +} - \sqrt{\tfrac{1}{1+\tau}} \, \ket{-\downarrow} \\
  V_0:      && \tfrac{\sqrt{\tau}}{1+\tau} \Bigbrk{ \ket{\uparrow\downarrow} - \ket{\downarrow\uparrow} }
              +\tfrac{1}{1+\tau}        \Bigbrk{ \tau \ket{++} + \ket{--} }
\ee
the ones in the second $\mathcal{R}_0$ also with $\Delta E = 0$ are
\be
  V_0:      && -\tfrac{\sqrt{\tau\vphantom{1}}}{1+\tau} \Bigbrk{ \ket{\uparrow\downarrow} - \ket{\downarrow\uparrow} }
               +\tfrac{1}{1+\tau} \Bigbrk{ \ket{++} + \tau \ket{--} } \\
  V_{1/2}:  && \sqrt{\tfrac{\tau\vphantom{1}}{1+\tau}} \, \ket{\uparrow -}   - \sqrt{\tfrac{1}{1+\tau}} \, \ket{+\uparrow}
        \comma \sqrt{\tfrac{\tau\vphantom{1}}{1+\tau}} \, \ket{\downarrow -} - \sqrt{\tfrac{1}{1+\tau}} \, \ket{+\downarrow} \\
  V_0:      && \ket{+-} \vphantom{\Bigbrk{}}
\ee
and the ones in $\mathcal{R}_{1/2}$ with $\Delta E = \frac{\mu\alpha^4}{2n^3}$ are
\be
  V_{1/2}:  &&  \sqrt{\tfrac{1}{1+\tau}} \, \ket{\uparrow +}   + \sqrt{\tfrac{\tau\vphantom{1}}{1+\tau}} \, \ket{-\uparrow}
        \comma  \sqrt{\tfrac{1}{1+\tau}} \, \ket{\downarrow +} + \sqrt{\tfrac{\tau\vphantom{1}}{1+\tau}} \, \ket{-\downarrow} \\
  V_0:      &&  -\tfrac{1-\tau}{1+\tau} \tfrac{1}{\sqrt{2}} \Bigbrk{ \ket{\uparrow\downarrow} - \ket{\downarrow\uparrow} }
                -\tfrac{2\sqrt{\tau}}{1+\tau} \tfrac{1}{\sqrt{2}} \Bigbrk{ \ket{++} - \ket{--} } \\
  V_1:      &&   \ket{\uparrow\uparrow}
        \comma   \tfrac{1}{\sqrt{2}} \Bigbrk{ \ket{\uparrow\downarrow} + \ket{\downarrow\uparrow} }
        \comma   \ket{\downarrow\downarrow} \\
  V_{1/2}:  &&   \sqrt{\tfrac{1}{1+\tau}} \, \ket{\uparrow -}   + \sqrt{\tfrac{\tau\vphantom{1}}{1+\tau}} \, \ket{+\uparrow}
        \comma   \sqrt{\tfrac{1}{1+\tau}} \, \ket{\downarrow -} + \sqrt{\tfrac{\tau\vphantom{1}}{1+\tau}} \, \ket{+\downarrow}
\ee

\subsection{$l>0$-states}

The relevant mixing matrices are \eqref{eqn:mixing-b-l>0} and \eqref{eqn:mixing-f-l>0}. They correspond to the subsector of states with fixed principal quantum number $n$, fixed orbital angular momentum $l$, and fixed $z$-component of the total angular momentum $m_j$. As we argued in \secref{sec:mixing}, there is no mixing with other subsectors even though sectors with different $l$ are degenerate at order $\alpha^2$.

Super multiplets, however, can only be formed by grouping together states with all possible values for $m_j$. The reason for this is that although a sector with fixed $(n,l,m_j)$ is closed under the action of the Hamiltonian, it is \emph{not} closed the action of the angular momentum operator $\vec{J}$ nor the supercharges $Q^\alpha$. These latter generators carry spin themselves, and therefore can change the $m_j$-value of the state they act on. Thus, we look at all $4\cdot4\cdot(2l+1)$ states with a given $l$-value at once. We find that they form two super multiplets $\mathcal{R}_{l}$ with $\Delta E=0$, one super multiplet $\mathcal{R}_{l+1/2}$ with $\Delta E = +\frac{\mu \alpha^4}{2(l+1)(2l+1) n^3}$, and one super multiplet $\mathcal{R}_{l-1/2}$ with $\Delta E = -\frac{\mu \alpha^4}{2l(2l+1) n^3}$.

The states of the first unperturbed $\mathcal{R}_{l}$ are given by\footnote{As discussed in \protect\secref{sec:mixing-l>0}, the states are labeled by $m_l = -l-1,-l,\ldots,l+1$, and kets that end up with a magnetic quantum number outside the interval $[-l,l]$ are defined to vanish.}
\be
 V_{l-\half}: &&
      \tfrac{1}{\sqrt{2l+1}} \Bigsbrk{
           \sqrt{l-m_l}   \Bigbrk{ \sqrt{\tfrac{\tau\vphantom{1}}{1+\tau}} \, \ket{m_l, \uparrow+}     - \sqrt{\tfrac{1}{1+\tau}} \, \ket{m_l, -\uparrow} } \nl \hspace{11mm}
         - \sqrt{l+m_l+1} \Bigbrk{ \sqrt{\tfrac{\tau\vphantom{1}}{1+\tau}} \, \ket{m_l+1, \downarrow+} - \sqrt{\tfrac{1}{1+\tau}} \, \ket{m_l+1, -\downarrow} }
      } \\[3mm]
 V_l:
  && \ket{m_l,-+} \vphantom{\Bigbrk{}} \\
 V_l:
  && \tfrac{\sqrt{\tau}}{1+\tau} \Bigbrk{ \ket{m_l,\uparrow\downarrow} - \ket{m_l,\downarrow\uparrow} }
     + \tfrac{1}{1+\tau} \Bigbrk{ \tau \ket{m_l, ++} + \ket{m_l, --} } \\[3mm]
 V_{l+\half}: &&
      \tfrac{1}{\sqrt{2l+1}} \Bigsbrk{
           \sqrt{l+m_l+1} \Bigbrk{ \sqrt{\tfrac{\tau\vphantom{1}}{1+\tau}} \, \ket{m_l, \uparrow+}     - \sqrt{\tfrac{1}{1+\tau}} \, \ket{m_l, -\uparrow} } \nl \hspace{11mm}
         + \sqrt{l-m_l}   \Bigbrk{ \sqrt{\tfrac{\tau\vphantom{1}}{1+\tau}} \, \ket{m_l+1, \downarrow+} - \sqrt{\tfrac{1}{1+\tau}} \, \ket{m_l+1, -\downarrow} }
      }
\ee
The ones in the second are
\be
 V_{l-\half}: &&
      \tfrac{1}{\sqrt{2l+1}} \Bigsbrk{
           \sqrt{l-m_l}   \Bigbrk{ \sqrt{\tfrac{\tau\vphantom{1}}{1+\tau}} \, \ket{m_l, \uparrow-}     - \sqrt{\tfrac{1}{1+\tau}} \, \ket{m_l, +\uparrow} } \nl \hspace{11mm}
         - \sqrt{l+m_l+1} \Bigbrk{ \sqrt{\tfrac{\tau\vphantom{1}}{1+\tau}} \, \ket{m_l+1, \downarrow-} - \sqrt{\tfrac{1}{1+\tau}} \, \ket{m_l+1, +\downarrow} }
      } \\[3mm]
 V_l:
  && \ket{m_l,+-} \vphantom{\Bigbrk{}} \\
 V_l:
  && -\tfrac{\sqrt{\tau}}{1+\tau} \Bigbrk{ \ket{m_l,\uparrow\downarrow} - \ket{m_l,\downarrow\uparrow} }
     + \tfrac{1}{1+\tau} \Bigbrk{ \ket{m_l, ++} + \tau \ket{m_l, --} } \\[3mm]
 V_{l+\half}: &&
      \tfrac{1}{\sqrt{2l+1}} \Bigsbrk{
           \sqrt{l+m_l+1} \Bigbrk{ \sqrt{\tfrac{\tau\vphantom{1}}{1+\tau}} \, \ket{m_l, \uparrow-}     - \sqrt{\tfrac{1}{1+\tau}} \, \ket{m_l, +\uparrow} } \nl \hspace{11mm}
         + \sqrt{l-m_l}   \Bigbrk{ \sqrt{\tfrac{\tau\vphantom{1}}{1+\tau}} \, \ket{m_l+1, \downarrow-} - \sqrt{\tfrac{1}{1+\tau}} \, \ket{m_l+1, +\downarrow} }
      }
\ee
The states in $\mathcal{R}_{l+1/2}$ which receive a positive energy shift by $\Delta E = +\frac{\mu \alpha^4}{2(l+1)(2l+1) n^3}$ are
\be
 V_l:  &&
     \tfrac{1}{\sqrt{2(l+1)(2l+1)}} \Bigsbrk{
          \sqrt{(l+m_l)(l-m_l+1)} \, \ket{m_l-1, \uparrow\uparrow} \nl \hspace{18mm}
        - \sqrt{(l-m_l)(l+m_l+1)} \, \ket{m_l+1, \downarrow\downarrow} \nl \hspace{18mm}
        - \tfrac{(l+m_l+1)-(l-m_l+1)\tau}{1+\tau} \, \ket{m_l, \uparrow\downarrow}
        + \tfrac{(l-m_l+1)-(l+m_l+1)\tau}{1+\tau} \, \ket{m_l, \downarrow\uparrow} \nl \hspace{18mm}
        - \tfrac{2(l+1)\sqrt{\tau}}{1+\tau} \Bigbrk{ \ket{m_l, ++} - \ket{m_l, --} }
      } \\[3mm]
 2V_{l+\half}:  &&
      \tfrac{1}{\sqrt{2l+1}} \Bigsbrk{
          \sqrt{l+m_l+1} \Bigbrk{ \sqrt{\tfrac{1}{1+\tau}} \, \ket{m_l, \uparrow\pm} + \sqrt{\tfrac{\tau\vphantom{1}}{1+\tau}} \, \ket{m_l, \mp\uparrow} } \nl \hspace{11mm}
        + \sqrt{l-m_l} \Bigbrk{ \sqrt{\tfrac{1}{1+\tau}} \, \ket{m_l+1, \downarrow\pm} + \sqrt{\tfrac{\tau\vphantom{1}}{1+\tau}} \, \ket{m_l+1, \mp\downarrow} }
      } \\[3mm]
 V_{l+1}: && 
      \tfrac{1}{\sqrt{2(l+1)(2l+1)}} \Bigsbrk{
           \sqrt{(l+m_l)(l+m_l+1)} \, \ket{m_l-1, \uparrow\uparrow} \nl \hspace{23mm}
         + \sqrt{(l-m_l)(l-m_l+1)} \, \ket{m_l+1, \downarrow\downarrow} \nl \hspace{23mm}
         + \sqrt{(l+m_l+1)(l-m_l+1)} \Bigbrk{ \ket{m_l, \downarrow\uparrow} + \ket{m_l, \uparrow\downarrow} }
      }
\ee
and the ones in $\mathcal{R}_{l-1/2}$ whose energy is lowered by $\Delta E = -\frac{\mu \alpha^4}{2l(2l+1) n^3}$ have the form
\be
 V_{l-1}: && 
      \tfrac{1}{\sqrt{2l(2l+1)}} \Bigsbrk{
           \sqrt{(l-m_l)(l-m_l+1)} \, \ket{m_l-1, \uparrow\uparrow} \nl \hspace{20mm}
         + \sqrt{(l+m_l)(l+m_l+1)} \, \ket{m_l+1, \downarrow\downarrow} \nl \hspace{20mm}
         - \sqrt{(l+m_l)(l-m_l)} \Bigbrk{ \ket{m_l, \downarrow\uparrow} + \ket{m_l, \uparrow\downarrow} }
      } \\[3mm]
 2V_{l-\half}:  &&
      \tfrac{1}{\sqrt{2l+1}} \Bigsbrk{
          \sqrt{l-m_l} \Bigbrk{ \sqrt{\tfrac{1}{1+\tau}} \, \ket{m_l, \uparrow\pm} + \sqrt{\tfrac{\tau\vphantom{1}}{1+\tau}} \, \ket{m_l, \mp\uparrow} } \nl \hspace{11mm}
        - \sqrt{l+m_l+1} \Bigbrk{ \sqrt{\tfrac{1}{1+\tau}} \, \ket{m_l+1, \downarrow\pm} + \sqrt{\tfrac{\tau\vphantom{1}}{1+\tau}} \, \ket{m_l+1, \mp\downarrow} }
      } \\[3mm]
 V_l:  &&
     \tfrac{1}{\sqrt{2l(2l+1)}} \Bigsbrk{
          \sqrt{(l+m_l)(l-m_l+1)} \, \ket{m_l-1, \uparrow\uparrow} \nl \hspace{15mm}
        - \sqrt{(l-m_l)(l+m_l+1)} \, \ket{m_l+1, \downarrow\downarrow} \nl \hspace{15mm}
        + \tfrac{(l-m_l)-(l+m_l)\tau}{1+\tau} \, \ket{m_l, \uparrow\downarrow}
        - \tfrac{(l+m_l)-(l-m_l)\tau}{1+\tau} \, \ket{m_l, \downarrow\uparrow} \nl \hspace{15mm}
        + \tfrac{2l\sqrt{\tau}}{1+\tau} \Bigbrk{ \ket{m_l, ++} - \ket{m_l, --} }
      }
\ee

\section{Conclusions and outlook}
\label{sec:conclusions}

A comprehensive summary of the results of our computation is given at the end of the introduction in \secref{sec:results}. Here we discuss some consequences and applications thereof.

\paragraph{Oscillations} Because of the energy splitting, there is an oscillation between different ``flavors''. Say we prepare an atom in the flavor state $\ket{++}$ with $l=0$, then it can oscillate into $\ket{--}$ and $\frac{1}{\sqrt{2}}\bigbrk{\ket{\uparrow\downarrow}-\ket{\downarrow\uparrow}}$. The probabilities of finding the atom in one of these states at time $t$ after the atom was purely $\ket{++}$ are
\be
 P_{\ket{++}} \eq \frac{1+6\tau^2+\tau^4+4\tau(1+\tau^2)\cos{\Delta E\, t}}{(1+\tau)^4} \\
 P_{\ket{--}} \eq \frac{8\tau^2(1-\cos{\Delta E\, t})}{(1+\tau)^4} \\
 P_{\ket{\uparrow\downarrow-\downarrow\uparrow}} \eq \frac{4\tau(1-\tau)^2(1-\cos{\Delta E t})}{(1+\tau)^4}
\ee
where $\Delta E = \frac{\mu\alpha^4}{2n^3}$. If we plug in the actual mass of the electron $m \approx 0.5\, \mathrm{MeV}$ and the muon $M \approx 100\, \mathrm{MeV}$, then $\Delta E \approx 0.7\, \mathrm{meV}$ for the lowest state ($n=1$). In the limit $m \ll M$, the oscillations have a frequency of 
\[
\omega \approx 10^{12}\, \mathrm{Hz}  \left( \frac{m}{500 \mbox{ keV}} \right) (137 \cdot \alpha)^4 \ .
\]

\paragraph{Supersymmetry} We can write the mixing matrices in terms of a superpotential $W$ as $M_b = W^\dagger W$ and $M_f = W W^\dagger$. The matrix $W$ can easily be constructed from the eigenstates given in \secref{sec:eigenstates} as follows. Let $\Lambda = \diag(E_1,\ldots,E_8)$ be the eigenvalues of $M_b$ and $M_f$ in some fixed order, and let $V$ and $U$ be matrices whose columns are the corresponding eigenvectors. Then the superpotential is given by $W = U \sqrt{\Lambda} V^\dagger$.  If we pair up the eigenvectors in $U$ and $V$ appropriately, we can set $W = Q_1 + Q_1^\dagger$ or $W=Q_2 + Q_2^\dagger$.  The prerequisite that all $E_i\ge0$ is satisfied for the $l=0$ sector, and in the other sectors we can achieve this requirement by adding the identity matrix times the smallest eigenvalue to the mixing matrices.

\paragraph{Bose condensate}

Ignoring interactions between these hydrogenic atoms, including any instability to form molecules, 
what happens if we place a large number of these atoms in a box?  If the atoms are fermionic, then only one fermionic atom can rest in the single particle ground state of the box.  (More generally, a small but finite number of fermionic atoms can rest in the ground state if the ground state has a small but finite degeneracy.)  In contrast, there is no limit to the number of bosonic atoms that can exist in the single particle ground state.  Through emission of a photino, a fermionic atom can convert into a bosonic atom.  Given our assumptions about the absence of interactions, the multiparticle ground state will contain at most one fermionic atom.  There should be no Fermi sea for these ``perfect atoms''.  It would be interesting to see what changes if any occur to this qualitative picture when interactions between the atoms are considered.

\paragraph{Supersymmetric chemistry}
For bound states of higher charge nuclei and more than one electron, the Pauli exclusion principle will play a much weaker role then it does in traditional atomic physics.  An electron in an excited orbital can always reduce its interaction energy with the nucleus by converting into a selectron and moving into a lower orbital at the possible price of increasing its interaction energy with other orbiting selectrons.  A supersymmetric periodic table should look quite different from the periodic table we are used to.  There may be additional interesting effects related to these atoms' ability to form molecules.  We leave a study of such effects for the future.

\section*{Acknowledgments}

We would like to thank Silviu Pufu, Stefan Stricker, Aleksi Vuorinen, and Lian-Tao Wang for discussions. We also thank Tomas Rube and Jay Wacker, who independently calculated the spectrum of supersymmetric hydrogen \cite{Wacker}, for bringing our attention to a mistake in an earlier version of this manuscript. The work of C.H. was supported in part by the US NSF under Grant Nos. PHY-0756966 and PHY-0844827.

\appendix

\section{Notation and conventions}
\label{app:conventions}

We use the metric $\eta^{\mu\nu} = \diag(-,+,+,+)$ and the Levi-Civita symbol $\levi^{12}=-\levi_{12}=1$. Implicit contractions of two-dimensional spinor indices are defined as $\psi\chi \equiv \psi^\alpha \chi_\alpha$, $\bar{\psi}\bar{\chi} \equiv \bar{\psi}_\dalpha \bar{\chi}^\dalpha$, and complex conjugation acts as $(\psi^\alpha)^\dagger = \bar{\psi}^\dalpha$, $(\psi_\alpha)^\dagger = \bar{\psi}_\dalpha$, $(\psi^\alpha\chi_\alpha)^\dagger = \bar{\chi}_\dalpha \bar{\psi}^\dalpha$. We raise and lower spinor indices from the left: $\psi^\alpha = \levi^{\alpha\beta} \psi_\beta$, $\psi_\alpha = \levi_{\alpha\beta} \psi^\beta$. We employ the Pauli matrices
\be
 \sigma^\mu_{\alpha\dalpha} = (-\unit, \vec{\sigma})
 \comma
 \bar{\sigma}^{\mu\dalpha\alpha} = \levi^{\dalpha\dbeta} \levi^{\alpha\beta} \sigma^\mu_{\beta\dbeta} = (-\unit, -\vec{\sigma})
\ee
to define the Dirac matrices as
\be
 \gamma^\mu = \matr{cc}{0 & \sigma^\mu \\ \bar{\sigma}^\mu & 0}
 \comma
 \gamma^5 = \gamma^0\gamma^1\gamma^2\gamma^3 = \matr{cc}{-i & 0 \\ 0 & i}
 \; .
\ee
The chiral projectors $P_\pm = \half(\unit\pm i\gamma^5)$ have the matrix representation $P_+ = \diag(1,0)$ and $P_- = \diag(0,1)$. The Clifford algebra relations are given by
$(\sigma^\mu \bar{\sigma}^\nu + \sigma^\nu \bar{\sigma}^\mu)_\alpha{}^\beta = - 2 \eta^{\mu\nu} \delta_\alpha{}^\beta$,
$(\bar{\sigma}^\mu \sigma^\nu + \bar{\sigma}^\nu \sigma^\mu)^\dalpha{}_\dbeta = - 2 \eta^{\mu\nu} \delta^\dalpha{}_\dbeta$,
$\acomm{\gamma^\mu}{\gamma^\nu} = - 2 \eta^{\mu\nu} \unit$.

\paragraph{Superfields and components} The vector superfield
$\VV = - \theta\sigma^\mu\bar{\theta} \, A_\mu(x)
       + i \, \theta^2 \, \bar{\theta} \bar{\chi}(x)
       - i \, \bar{\theta}^2 \, \theta \chi(x)
       + \half \theta^2 \, \bar{\theta}^2 \, \auxD(x)$
contains the photon $A_\mu$, the gaugino $\chi$, and the auxiliary real scalar $\auxD$. The chiral superfields
$\Phi_\pm = \phi_\pm(y) + \sqrt{2} \, \theta\psi_\pm(y) + \theta^2 \, F_\pm(y)$ 
(where $y^\mu = x^\mu + i \theta \sigma^\mu \bar{\theta}$) 
contain the slepton $\phi_\pm$, the leptons $\psi_\pm$, and the auxiliary complex scalars $F_\pm$. We introduce a Dirac spinor for the leptons
\be
 \psi = \matr{c}{ \psi_{+\alpha} \\ \bar{\psi}_-^\dalpha }
 \comma
 \bar{\psi} = \psi^\dagger \gamma^0 = \matr{cc}{ -\psi_-^\alpha & -\bar{\psi}_{+\dalpha} } \; ,
\ee
and a Majorana spinor for the photino
\be
 \lambda = \matr{c}{ \chi_\alpha \\ \bar{\chi}^\dalpha }
 \comma
 \bar{\lambda} = \lambda^\dagger \gamma^0 = \matr{cc}{ -\chi^\alpha & -\bar{\chi}_\dalpha } \; ,
\ee
subject to the condition $\lambda = \lambda^C \equiv C \bar{\lambda}^\trans$ with the charge conjugation matrix $C = i\gamma^0\gamma^2 = - C^\trans = - C^\dagger = C^* = -C^{-1} = \diag(i\sigma^2, -i\sigma^2)$. The sign in the gauge covariant derivative $\deriD_\mu X = \partial_\mu X + iq A_\mu X$ is determined by the $\grU(1)$ charge $q$ of $X$. The field strength is $F_{\mu\nu} = \partial_\mu A_\nu - \partial_\nu A_\mu$.

\section{Bound states in Coulomb potential}
\label{sec:Coulomb-problem}

The free Hamiltonian is given by $H_0 = \frac{\vec{p}^{\,2}}{2\mu} - \frac{\alpha}{r}$. We use the constants $\mu = \frac{mM}{m+M}$ for the reduced mass, $\alpha = \frac{e^2}{4\pi}$ for the fine structure constant, and $a_B = \frac{1}{\mu \alpha}$ for the Bohr radius. The wave functions $\psi_{nlm}(\vec{r}) = \braket{\vec{r}}{nlm}$ for the bound states are given by
\be
  \psi_{nlm}(\vec{r}) = \frac{1}{\sqrt{a_B^3}} \frac{2}{n^2} \sqrt{\frac{(n-l-1)!}{(n+l)!}} \lrbrk{\frac{2r}{n a_B}}^l L_{n-l-1}^{2l+1}\Bigbrk{\frac{2r}{n a_B}} \exp \Bigbrk{-\frac{r}{n a_B}} Y_{lm}(\theta,\varphi) \; .
\ee
Their energies are $E_n = - \frac{\mu \alpha^2}{2n^2}$. In Mathematica one has to write $L_a^b(x) = \mathtt{LaguerreL[a,b,x]}$ and $Y_{lm}(\theta,\varphi) = \mathtt{SphericalHarmonicY[l,m,\theta,\varphi]}$.
\paragraph{Expectation values} 
\be
  \bra{nlm}\frac{1}{r}\ket{nlm} = \frac{\mu \alpha}{n^2}  \; , \; \; \;
 \bra{nlm}\frac{1}{r^2} \ket{nlm} = \frac{ \mu^2 \alpha^2}{n^3 (l + \frac{1}{2})} \; , \; \; \;
  \bra{n00}\frac{r_i r_j}{r^3}\ket{n00} = \frac{\mu\alpha\delta_{ij}}{3n^2} 
  \ee
 \be
  \bra{nlm}\frac{\vec{p}\,{}^2}{r}\ket{nlm} = \frac{\mu^3\alpha^3}{n^4} \lrsbrk{\frac{2n}{l+\half}-1} 
 \; , \; \; \;
   \bra{nlm}\vec{p}\,{}^4\ket{nlm} = \frac{\mu^4\alpha^4}{n^4} \lrsbrk{\frac{4n}{l+\half}-3} 
 \ee
 \begin{eqnarray}
  \bra{n'l'm'} \delta^{(3)}(\vec{r}) \ket{nlm} &=& \frac{\mu^3 \alpha^3}{\pi n^3} \delta_{nn'}\delta_{l0}\delta_{l'0}\delta_{m0}\delta_{m'0} \\
  \bra{nl'm'} \frac{1}{r^3} \ket{nlm} &=& \frac{\mu^3 \alpha^3}{n^3 l(l+\half)(l+1)} \delta_{ll'}\delta_{mm'} \qquad \mbox{for $l,l' > 0$}\\
  \bra{nl'm'} \frac{ (\vec{r}\cdot\vec{p})^2}{r^3} + 2 \pi \delta^{(3)}(\vec r) \ket{nlm}
  &=&
  \frac{\mu^3 \alpha^3}{n^4} \left[ \frac{n}{l + \frac{1}{2}} - 1 \right] \delta_{ll'}\delta_{mm'}
\end{eqnarray}
\paragraph{Angular integrals} Define components of the unit position vector $\hat{\vec{r}}$ as $\hat{r}_0 \equiv \frac{z}{r} = \cos\theta$ and $\hat{r}_\pm \equiv \frac{x\pm iy}{r} = \sin\theta e^{\pm i\varphi}$.
\begin{flalign}
  &\bra{l'm'} \hat{r}_0 \ket{lm}
    =  \sqrt{\frac{(l'+m')(l'-m')}{(2l'+1)(2l+1)}} \delta_{l',l+1} \delta_{m',m} 
      + \sqrt{\frac{(l+m)(l-m)}{(2l'+1)(2l+1)}} \delta_{l',l-1} \delta_{m',m} \\
  &\bra{l'm'} \hat{r}_\pm \ket{lm}
    = \mp \sqrt{\frac{(l'\pm m'-1)(l'\pm m')}{(2l'+1)(2l+1)}} \delta_{l',l+1} \delta_{m',m\pm 1} 
      \pm \sqrt{\frac{(l\mp m-1)(l\mp m)}{(2l'+1)(2l+1)}} \delta_{l',l-1} \delta_{m',m\pm 1} 
\end{flalign}
\be
 \label{r0r0}
  \bra{l m} \hat{r}_0 \hat{r}_0 \ket{l m} = \frac{2l(l+1) - 2m^2 -1}{(2l+3)(2l-1)} \; , \; \; \;
  \bra{l m} \hat{r}_+ \hat{r}_- \ket{l m} = \frac{2l(l+1) + 2m^2 -2}{(2l+3)(2l-1)} \; ,
\ee
\be
  \bra{l \, m+1} \hat{r}_+ \hat{r}_0 \ket{l m } = -\frac{(2m+1) c_{lm}}{(2l+3)(2l-1)} \; , \; \; \;
  \bra{l \, m+1} \hat{r}_+ \hat{r}_+ \ket{l \, m-1} = - \frac{2c_{lm}c_{l,-m}}{(2l+3)(2l-1)} \; ,
  \label{rprp}
\ee
where here and below we use
\be
  c_{lm} \equiv \sqrt{(l-m)(l+m+1)} \; .
\ee

\section{Feynman Rules}

\begin{samepage}\paragraph{Fields and particles}
\begin{center}
\renewcommand{\arraystretch}{1.4}
\begin{tabular}{l|cccc|cccc|cc}
Field       & $\phi_+$  & $\phi_+^\dagger$ & $\phi_-$  & $\phi_-^\dagger$ & $\psi_+$ & $\bar{\psi}_+$ & $\psi_-$ & $\bar{\psi}_-$ & $\chi$             & $\bar{\chi}$       \\ \hline
creates     & $\tilde{e}^+_+$ & $\tilde{e}^-_+$        & $\tilde{e}^-_-$ & $\tilde{e}^+_-$        & $e^+_+$  & $e^-_+$        & $e^-_-$  & $e^+_-$        & $\tilde{\gamma}_+$ & $\tilde{\gamma}_-$ \\ \hline
annihilates & $\tilde{e}^-_+$ & $\tilde{e}^+_+$        & $\tilde{e}^+_-$ & $\tilde{e}^-_-$        & $e^-_+$  & $e^+_+$        & $e^+_-$  & $e^-_-$        & $\tilde{\gamma}_-$ & $\tilde{\gamma}_+$ \end{tabular}
\renewcommand{\arraystretch}{1}
\end{center}
The $\grU(1)_e$ charges of the particles are encoded in the labels as $Q(\tilde{e}^q_h) = q$, $Q(e^q_h) = q$, $Q(\tilde{\gamma}_h) = 0$, and the $\grU(1)_R$ are given by $R(\tilde{e}^q_h) = q h$, $R(e^q_h) = 0$, $R(\tilde{\gamma}_h) = h$.
\end{samepage}

\paragraph{Propagators} 
The momentum $p$ flows from $y$ to $x$. For photons in Coulomb gauge
\be
  \vev{A_\mu(x)A_\nu(y)} \rightarrow
 \tilde{\Delta}_{00}(p) = \frac{i}{\vec p \, {}^2} \ , \; \;
 \tilde{\Delta}_{0i}(p) = 0 \ , \; \; \;
  \tilde{\Delta}_{ij}(p) =
- \frac{i}{p^2} \left( \delta_{ij} - \frac{p_i p_j}{\vec p \, {}^2} \right)  \ .
\ee
Coulomb gauge is better suited for taking a non-relativistic limit of the SQED amplitudes. For a discussion of the complications that arise when combining Feynman gauge with a non-relativistic limit, see for example \cite{Lindgren}. For photinos
\be
  \vev{\lambda_\alpha(x)\bar{\lambda}_\beta(y)} \rightarrow
  \tilde{\Delta}_{\alpha\beta}(p) =
  \lrbrk{\frac{-i}{\slashed{p}+i\eps}}_{\alpha\beta} =
  \frac{i\slashed{p}_{\alpha\beta}}{p^2-i\eps} \ .
\ee

\paragraph{Electron wave functions}
\be
 \psi(x) = \int\frac{d^3p}{(2\pi)^3} \frac{1}{\sqrt{2E_{\vec{p}}}} \Bigsbrk{ b^s_{\vec{p}} u^s(p) e^{ipx} + d^{s\dagger}_{\vec{p}} v^s(p) e^{-ipx} }
\ee
Contractions with external particles are given by
\begin{align}
  \bra{e^-(p,s)} \bar{\psi} &= \bar{u}^s(p) \, e^{-ipx} &
  \psi \ket{e^-(p,s)}       &= u^s(p) \, e^{ipx} \\
  \bra{e^+(p,s)} \psi       &= v^s(p) \, e^{-ipx} &
  \bar{\psi} \ket{e^+(p,s)} &= \bar{v}^s(p) \, e^{ipx}
\end{align}
It is useful to know the identities
\be
 \sqrt{\sigma\cdot p} = \frac{\sigma\cdot p+m}{\sqrt{2(p^0+m)}}
 \comma
 \sqrt{\bar{\sigma}\cdot p} = \frac{\bar{\sigma}\cdot p+m}{\sqrt{2(p^0+m)}}
 \; .
\ee
In the quantum mechanical setting we convert the spinor $\eta$ for an anti-particle into a spinor for a particle using the relation $\eta = C \xi^*$ where the charge conjugation matrix is $C=i\sigma^2$.

\section{Degenerate Perturbation Theory}

\subsection{$l=0$}

The only non-zero matrix elements of the interaction Hamiltonian 
\be
H_{\mathrm{deg}} = H_{\mathrm{int}} + \frac{\vec p \, {}^4}{8} \left( \frac{1}{m^3} + \frac{1}{M^3} \right) + \frac{e^2}{4 \pi} \frac{\vec p \, {}^2}{M m r} \ ,
\ee
for s-wave states $\ket{n00\,s_es_m}$ are 
\be
  \bra{n00\,{\xi_{eo}\xi_{mo}}} H_{\mathrm{deg}} \ket{n00\,{\xi_{ei}\xi_{mi}}}
    \eq \frac{\mu^3 \alpha^4}{n^3} \, \xi^\dagger_{mo} \xi^\dagger_{eo} \biggsbrk{ \frac{1}{2} \biggbrk{\frac{1}{m}+\frac{1}{M}}^2 + \frac{2}{3Mm} \vec{\sigma}_e \cdot \vec{\sigma}_m } \xi_{ei} \xi_{mi} \ , \\
  \bra{n00\,{\pm\pm}} H_{\mathrm{deg}} \ket{n00\,{\xi_{ei}\xi_{mi}}}
    \eq \mp \frac{1}{2}  \frac{\mu^3 \alpha^4}{n^3}  \frac{M-m}{(Mm)^{3/2}}  \, \xi^\trans_{mi} \,i\sigma^2 \xi_{ei} \ ,  \\
  \bra{n00\,{\xi_{eo}\xi_{mo}}} H_{\mathrm{deg}} \ket{n00\,{\pm\pm}}
    \eq \pm  \frac{1}{2} \frac{\mu^3 \alpha^4}{n^3}  \frac{M-m}{(Mm)^{3/2}}   \, \xi^\dagger_{eo} i\sigma^2\, \xi^*_{mo}\ ,  \\
  \bra{n00\,{\pm\pm}} H_{\mathrm{deg}} \ket{n00\,{\pm\pm}}
    \eq \frac{\mu^3 \alpha^4}{n^3} \; \frac{2}{Mm}
\ee
for bosonic atoms, and
\be
  \bra{n00\,{\xi_{eo}\pm}} H_{\mathrm{deg}} \ket{n00\,{\xi_{ei}\pm}}
    \eq \frac{\mu^3 \alpha^4}{n^3} \left( \frac{1}{Mm}+\frac{1}{2m^2} \right) \, \xi^\dagger_{eo} \xi_{ei} \ , \\
  \bra{n00\,{\pm\xi_{mo}}} H_{\mathrm{deg}} \ket{n00\,{\pm\xi_{mi}}}
    \eq \frac{\mu^3 \alpha^4}{n^3} \left( \frac{1}{Mm}+\frac{1}{2M^2} \right) \, \xi^\dagger_{mo} \xi_{mi} \ , \\
  \bra{n00\,{\mp\xi_{mo}}} H_{\mathrm{deg}} \ket{n00\,{\xi_{ei}\pm}}
    \eq  \frac{1}{2} \frac{\mu^3 \alpha^4}{n^3}  \frac{M+m}{(Mm)^{3/2}}  \, \xi^\dagger_{mo} \xi_{ei} \ , \\
  \bra{n00\,{\xi_{eo}\mp}} H_{\mathrm{deg}} \ket{n00\,{\pm\xi_{mi}}}
    \eq \frac{1}{2}  \frac{\mu^3 \alpha^4}{n^3}\frac{M+m}{(Mm)^{3/2}}   \, \xi^\dagger_{eo} \xi_{mi}
\ee
for fermionic ones.

\subsection{$l>0$}

States of different $n$ and $l$ do not mix,
and we fix $n$ and $l$.
We begin with the bosonic states (\ref{bosonicin}). 
The only non-zero matrix elements of the interaction Hamiltonian are
\be
\bra{nl m_l' \,{\xi_{eo}\xi_{mo}}} H_{\mathrm{deg}} \ket{nlm_l \,{\xi_{ei}\xi_{mi}}}
\eq 
   \frac{C}{2}    \left \langle 
    l m_l' \left| \xi^\dagger_{mo} \xi^\dagger_{eo}  \left(
   \vec L \cdot \vec \sigma_e \left( \frac{1}{2m^2} + \frac{1}{Mm} \right) +
   \right . \right . \right. \\
   &&
   + \vec L \cdot \vec \sigma_m \left( \frac{1}{2M^2} + \frac{1}{Mm} \right) + \nonumber \\
   &&
   \left. \left. \left.
   + \frac{1}{2 Mm} \left( 3 \hat r \cdot \vec \sigma_e \hat r \cdot \vec \sigma_m - \vec \sigma_e \cdot \vec \sigma_m \right)
   \right)   \xi_{ei} \xi_{mi} \right|l m_l  \right \rangle \ , \nonumber
    \\
  \bra{nl m_l' \,{\pm\pm}} H_{\mathrm{deg}} \ket{nl m_l \,{\xi_{ei}\xi_{mi}}}
    \eq 
   \mp  \frac{C}{4} \frac{M+m}{(Mm)^{3/2}} 
    \left \langle l m_l'\left |  
   \vec L   \right| l m_l  \right \rangle \cdot
    \left( \xi^\trans_{mi} \,i\sigma^2 \, \vec \sigma \xi_{ei} \right)
   \ , \\
  \bra{nl m_l' \,{\xi_{eo}\xi_{mo}}} H_{\mathrm{deg}} \ket{nl m_l\,{\pm\pm}}
    \eq
    \pm \frac{C}{4} \frac{M+m}{(Mm)^{3/2}}  \left \langle l m_l' \left | \vec L  \right| l m_l \right \rangle
    \cdot 
    \left( \xi^\dagger_{eo}  \vec \sigma \, i\sigma^2\, \xi^*_{mo} \right)\ , 
\ee
where
\be
C \equiv \frac{\mu^3 \alpha^4}{n^3 l (l+\frac{1}{2}) (l+1)} \ .
\ee
The nonzero matrix elements of the interaction Hamiltonian for the fermionic states (\ref{fermionicin}) are
\be
  \bra{nl m_l' \,{\xi_{eo}\pm}} H_{\mathrm{deg}} \ket{nl m_l \,{\xi_{ei}\pm}}
    \eq 
    \frac{C}{2} \left( \frac{1}{Mm}+\frac{1}{2m^2} \right)  \, 
    \langle l m_l' \, {\xi_{eo} \pm} | \vec L \cdot \vec \sigma | l m_l \, \xi_{ei} \pm  \rangle
    \ , \\
  \bra{nl m_l' \,{\pm\xi_{mo}}} H_{\mathrm{deg}} \ket{nl m_l \,{\pm\xi_{mi}}}
  \eq
   \frac{C}{2} \left( \frac{1}{Mm}+\frac{1}{2M^2} \right)  \, 
   \langle l m_l' \, {\pm \xi_{mo}} | \vec L \cdot \vec \sigma | l m_l \, {\pm \xi_{mi}} \rangle \ , \\
  \bra{nl m_l' \,{\mp\xi_{mo}}} H_{\mathrm{deg}} \ket{nl m_l \,{\xi_{ei}\pm}}
  \eq
    \frac{C}{4} \frac{M+m}{(Mm)^{3/2}}  \, \langle l m_l' | \vec L |
    l m_l   \rangle    \cdot  \xi_{mo}^\dagger \vec \sigma \xi_{ei} \ , \\
  \bra{nl m_l' \,{\xi_{eo}\mp}} H_{\mathrm{deg}} \ket{nl m_l \,{\pm\xi_{mi}}}
  \eq
    \frac{C}{4} \frac{M+m}{(Mm)^{3/2}}\, \langle l m_l'  | \vec L  | l m_l  \rangle 
   \cdot  \xi_{eo}^\dagger \vec \sigma \xi_{mi}
    \ .
\ee

Thus to determine the energy corrections from first 
order degenerate perturbation theory, 
we need the matrix elements of  $ \vec L \cdot \vec \sigma$,
$\vec \sigma_e \cdot \vec \sigma_m $,
and $ \hat r \cdot \vec \sigma_e \hat r \cdot \vec \sigma_m $:
\[
\begin{array}{r|cc}
\langle \vec L \cdot \vec \sigma \rangle &
    | m_l \, \uparrow \rangle &
    | m_l+1 \, \downarrow  \rangle \\
    \hline
\langle m_l \, \uparrow | 
& m_l
& c_{lm_l}
\\
\langle m_l+1 \, \downarrow  | 
& c_{lm_l}
& -(m_l+1) 
\end{array} \ ,
\]

\[
\begin{array}{r|cccc}
\langle \vec \sigma_e \cdot \vec \sigma_m \rangle & | m_l-1 \, \uparrow \uparrow \rangle &
    | m_l \, \downarrow \uparrow \rangle &
    | m_l \, \uparrow \downarrow  \rangle &
    | m_l+1 \, \downarrow \downarrow \rangle \\
\hline
\langle m_l-1 \, \uparrow \uparrow | &
1 
& 0
& 0
& 0
\\
\langle m_l \, \downarrow \uparrow | 
& 0
& -1 
& 2 
& 0
\\
\langle m_l \,  \uparrow \downarrow  | 
& 0
& 2
& -1
& 0
\\
\langle m_l+1 \, \downarrow \downarrow | 
& 0
& 0
& 0
& 1 
\end{array} \ ,
\]

\[
\begin{array}{r|cccc}
\langle \hat r \cdot \vec \sigma_e \hat r \cdot \vec \sigma_m \rangle & | m_l-1 \, \uparrow \uparrow \rangle &
    | m_l \, \downarrow \uparrow \rangle &
    | m_l \, \uparrow \downarrow \rangle &
    | m_l+1 \, \downarrow \downarrow \rangle \\
\hline
\langle m_l-1 \, \uparrow \uparrow | 
& \cos^2 \theta 
& \sin \theta \cos \theta \, e^{-i \phi}
& \sin \theta \cos \theta \, e^{-i \phi}
& \sin^2 \theta \, e^{-2 i \phi}
\\
\langle m_l \, \downarrow \uparrow | 
& \sin \theta \cos \theta \, e^{i \phi}
&  -\cos^2 \theta
&  \sin^2 \theta
&  -\sin \theta \cos \theta \, e^{-i \phi} 
\\
\langle m_l \, \uparrow \downarrow | 
& \sin \theta \cos \theta \, e^{i \phi}
& \sin^2 \theta
& -\cos^2 \theta
& - \sin \theta \cos \theta \, e^{-i \phi}
\\
\langle m_l+1 \, \downarrow \downarrow | 
& \sin^2 \theta \, e^{2 i \phi}
& -\sin \theta \cos \theta \, e^{i \phi}
& -\sin \theta \cos \theta \, e^{i \phi}
& \cos^2 \theta 
\end{array} \ .
\]
In evaluating this last matrix, the integrals (\ref{r0r0}--\ref{rprp}) are useful.
We also find
\[
\begin{array}{r|cccc}
\pm \langle i \sigma^2 \vec L \cdot \vec \sigma \rangle & | m_l-1 \, \uparrow \uparrow \rangle &
    | m_l \, \downarrow \uparrow \rangle &
    | m_l \, \uparrow \downarrow \rangle &
    | m_l+1 \, \downarrow \downarrow \rangle \\
\hline
\langle m_l \, + +| 
& c_{l,-m_l}
& -m_l
& -m_l
& - c_{lm_l}
\\
\langle m_l \, - -| 
& -c_{l,-m_l}
& m_l
& m_l
& c_{lm_l}
\end{array} \ .
\]

\section{Second Order Perturbation Theory}
\label{app:secorderpert}

We are interested in computing the correction to the energy of a state $| n l m, s_e s_m \rangle$ at second order in perturbation theory.  (For ease of notation, we remove the subscript $l$ from $m_l$ in this subsection and replace $m$ with $m_e$.)    These second order corrections will not mix states of different $n$ and $l$, $n$ because the energies are different and $l$ because of the vanishing of an integral we discuss below.  We need to compute the matrix 
\be
\Delta E(n,l,m,m', s_e, s_e', s_m, s_m') = {\sum_{i}}' \frac{ \langle n l m, s_e s_m | H_{\rm int} | i \rangle \langle i | H_{\rm int} | n l m', s_e' s_m' \rangle}{E_n - E_i} \ ,
\label{secorderv2}
\ee
where the $'$ on the sum means we should omit the states with $E_i = E_n$.  (We can omit these states in the sum because of eq.\ (\ref{intervanish}).)  This sum involves both the discrete and continuum hydrogenic states. 

The Hermitian matrix $\Delta E$ has a block diagonal form.  The four bosonic states
\[
 \ket{m-1 \uparrow \uparrow},
  \ket{m \downarrow \uparrow},
  \ket{m \uparrow \downarrow},
  \ket{m+1 \downarrow \downarrow}
  \]
  mix among themselves according to
 \be
 2 \matr{cccc}{
 \Delta E_{-+}(m-1) & -\Delta E_{-0}(m) & -\Delta E_{-0}(m) & -\Delta E_{--}(m+1) \\
 -\Delta E_{0+}(m-1) & \Delta E_{00}(m) & \Delta E_{00}(m) & \Delta E_{0-}(m+1)  \\
 -\Delta E_{0+}(m-1) & \Delta E_{00}(m) & \Delta E_{00}(m) & \Delta E_{0-}(m+1) \\
 -\Delta E_{++}(m-1) & \Delta E_{+0}(m) & \Delta E_{+0}(m) & \Delta E_{+-}(m+1)  
 } \ .
\ee
The factor of two comes from the sum over scalar intermediate states $++$ and $--$.  We will define $\Delta E_{ij}(m)$ presently.  The coefficients of $\Delta E$ vanish for the other four bosonic states
\[
 \ket{m, ++},
    \ket{m, --} ,
  \ket{m, +-},
  \ket{m, -+}
  \ ,
\]
provided $l>0$.  In the special case $l=0$, the states $\ket{m, ++}, \ket{m,--}$  mix as
\be
2 \matr{cc}{
\Delta E_r & \Delta E_r \\
\Delta E_r & \Delta E_r
}
\ee
where $\Delta E_r = \Delta E_{00} + (\Delta E_{+-} + \Delta E_{-+})/2$.
For the eight fermionic states, $\Delta E$ reduces to four $2\times 2$ blocks.  The states 
$\ket{m, \uparrow \pm}, \ket{ m+1, \downarrow \pm}$ and the states
$\ket{m, \pm \uparrow}, \ket{m+1, \pm \downarrow}$ each have the same second order mixing matrix
\be
 \left(
\begin{array}{cc}
\Delta E_{00}(m)+\Delta E_{-+}(m)  &\Delta E_{-0}(m+1) - \Delta E_{0-}(m+1)  \\
\Delta E_{0+}(m)-\Delta E_{+0}(m) & \Delta E_{00}(m+1) + \Delta E_{+-}(m+1)  \\
\end{array}
\right) \ .
\ee
The orbital part of the interaction that contributes at second order in 
perturbation theory has the schematic form $(p-q)_i / (\p - \q)^2$.  
We will work in a basis where $p_0 = p_z$ and $p_\pm = (p_x \pm i p_y)$.

 To compute the sum (\ref{secorderv2}), we make use of a beautiful result of Schwinger \cite{Schwinger:1964} for the Coulomb Green's function in momentum space:
 \be
 \left( E - \frac{\vec p \, {}^2}{2 \mu} \right) G(\p, \p \, {}'; E) 
  + \frac{\alpha}{2 \pi^2} \int \frac{ d^3 \p \, {}''}{(\p - \p \, {}'')^2} G(\p \, {}'', \p \, {}'; E) = \delta^{(3)} (\p - \p \, {}') \ .
 \ee
Introducing a $p_0$ such that $E = - p_0^2 / 2 \mu$, 
the construction makes use of a mapping of the four momentum $(p_0, \p)$ to an $S^3$:
 \be
 \vec \xi = - \frac{2 p_0}{\p \, {}^2 + p_0^2} \p \, , \; \; \; \xi_0 = \frac{ p_0^2 - \p \, {}^2}{p_0^2 + \p \, {}^2} \ ,
 \; \; \; 
 \vec \xi \, {}^2 + \xi_0^2 = 1 \ .
 \ee
 Let $\Omega = (\psi, \theta, \phi)$ be a set of angular coordinates on the $S^3$.
The matrix elements in the Hamiltonian at second order in perturbation theory can be built from the expression:
\begin{equation}
\label{DeltaEinit}
\Delta E_{ij}(m) = \frac{4 \pi^2 \alpha^2}{M m_e} \int \frac{ d^3 \p \, d^3 \p\,' \, d^3 \q \, d^3 \q\,'}{(2\pi)^6}
\phi_{n, l, m +i+j}^*(p) \frac{(p-p')_i}{(\p-\p\,')^2} G(p', q'; E) 
\frac{(q-q')_j}{(\q - \q\,')^2} \phi_{n l m} (q) \ .
\end{equation}
where
\begin{equation}
G(p', q'; E) = - \frac{16 \mu p_0^3}{(p_0^2 + \p\,'{}^2)^2 (p_0^2 + \q\,'{}^2)^2}
{\sum_{n' l' m'}} \frac{Y_{n'l' m'}(\Omega_{p'}) Y_{n' l' m'}^* (\Omega_{q'})}{1-\alpha \mu/ n' p_0} \ .
\end{equation}
The conserved energy is $E = - \mu \alpha^2 / 2 n^2$.
The $\phi_{nlm}$ are the hydrogen wave functions in momentum space:
\begin{equation}
\phi_{nl m}(p) 
=  \frac{4 (p_0)^{5/2}}{( \p\, {}^2 + p_0^2)^2} Y_{n l m} (\Omega) 
\; \; \; ; \; \; \; Y_{n l m} (\Omega)  =  Z_{nl}(\psi)Y_{l m} (\theta, \varphi) \ .
\end{equation}
The $Y_{nlm}(\Omega)$ are spherical harmonics on the momentum $S^3$.  The $Y_{nlm}$ and $Y_{lm}$ satisfy similar orthonormality and completeness relations.
More explicitly,
\begin{equation}
Z_{nl}(\psi) = N_{nl} \, \sin^l \psi \, C_{n-l-1}^{l+1} (\cos \psi) \ ,
\end{equation}
where the $C_n^l(x)$ are the Gegenbauer polynomials and
\[
N_{n l} = \left[ \frac{n(n- l-1)!}{(n+l)!} \frac{2^{2 l+1} (l!)^2}{\pi} \right]^{1/2} \ .
\]

Making use of the relation
\be
\frac{1}{4\pi^2} \frac{1}{(\xi - \xi')^2} = \sum_{nlm} \frac{1}{2n} Y_{nlm}(\Omega) Y_{nlm}^*(\Omega') \ ,
\ee
the expression (\ref{DeltaEinit}) simplifies to
\begin{equation}
\Delta E_{ij}(m) = - \frac{\mu^3 \alpha^4}{M m_e} \frac{1}{ n^2} {\sum_{n' l' m'}} \frac{n'}{n'-n} 
S_i^{n'l' m',m+i+j} {S_j^{n' l' m', m}}^* \ ,
\end{equation}
where
\begin{equation}
S_i^{n' l' m',m} = 
\frac{n'-n}{2nn'} \int d\Omega \, Y_{n l m}^*(\Omega) \frac{\xi_i}{1+ \xi_0}Y_{n' l' m'}(\Omega) \ .
\end{equation}
To perform this integral, we use the definition of the $Y_{n l m}(\Omega)$:
\begin{equation}
S_i^{n' l' m',m} = \frac{n'-n}{2nn'} \, \langle l m | \hat r_i | l' m' \rangle \, \int Z^*_{n l}(\psi) \frac{ \sin^3 \psi}{1+ \cos \psi} Z_{n' l'}(\psi) \, d\psi
 \ .
\end{equation}
The integral $\langle l m | \hat r_i | l' m' \rangle$ vanishes unless $l = l' \pm 1$ by the electric dipole selection rules.  The energy shift simplifies to
\begin{equation}
\Delta E_{ij}(m) = - \frac{ \mu^3 \alpha^4}{M m_e n^4} {\sum_{n' l'}} \left[ c_- \delta_{l', l-1}  
+c_+ \delta_{l', l+1}  \right] \frac{n'-n}{4 n'} 
\left|
\int Z_{n l}^*(\psi) \frac{\sin^3 \psi}{1+\cos \psi} Z_{n' l'}(\psi) \, d \psi
\right|^2 \ ,
\end{equation}
where $c_\pm$ depends on $i$, $j$, $l$, and $m$ but not on $n'$, $l'$, or $m'$.

Let's define
\begin{equation}
I(n, n', l, l') \equiv  \int_0^\pi \frac{(\sin \psi)^{l + l' +3}}{1+\cos \psi} C_{n'-l'-1}^{l'+1} (\cos \psi) C_{n - l-1}^{l+1} (\cos \psi) \, d \psi  \ .
\end{equation}
For positive integers $n$ and $n'$, we find that
\begin{equation}
| I(n,n', l, l+1)|^2 = 
\begin{cases}
\left( \frac{ \pi (n+ l)!}{2^{2l+1} l! (l+1)! (n-l-1)!} \right)^2 \ , & n < n'  \ , \\
0 \ , &  n> n' \ . 
\end{cases} 
\end{equation}
Given these results, we can evaluate the sum for $l>0$:
\begin{eqnarray}
\Delta E_{ij}(m) &=& - \frac{ \mu^3 \alpha^4}{M m_e n^4} \left[
c_-  \frac{ n (n-l-1)!}{(n+l)!} \sum_{n'=1}^{n-1} (n'-n) \frac{(n'+ l-1)!}{(n'-l)!}
\right.
\nonumber \\ 
&& 
\left.
\hspace{20mm} +c_+ \frac{ n(n+l)!}{(n-l-1)!} \sum_{n'=n+1}^\infty
(n'-n) \frac{(n'-l-2)!}{(n'+l+1)!} \right] \nn \\
&=& - \frac{ \mu^3 \alpha^4}{M m_e n^3} \left[ c_+ \frac{1}{(2l+1)(2l+2)} - c_- \frac{1}{2l(2l+1)} \right]\ .
\end{eqnarray}
(If $l=0$, the coefficient of $c_-$ will vanish because the state $l' = l-1$ does not exist.)

From the integral $I$, we deduce that there is no mixing between $l$ and $l+2$ states at second order in perturbation theory.  Note that $I(n,n',l,l') = I(n',n,l',l)$.  To get mixing between these states we need some amplitude to scatter from an $nl$ state to an $n', l+1$ state and back to a $n, l+2$ state.  In other words, the product
$I(n,n',l,l+1) I(n',n,l+1,l+2)$ should not vanish.  However, $I(n',n,l+1,l+2)$ will vanish  unless $n' < n$ while $I(n,n',l,l+1)$ will vanish unless $n<n'$.

To evaluate the $\Delta E_{ij}$ completely, we need
\be
 && \sum_{m'=-l'}^{l'} \bra{lm} \hat{r}_i \ket{l'm'} \bra{l'm'} \hat{r}_i \ket{lm} =
  \frac{1}{2l+1} \Bigsbrk{  l \, \delta_{l',l-1} + (l+1) \, \delta_{l',l+1} } \ , \\
 && \sum_{m'=-l'}^{l'} \bra{lm} \hat{z} \ket{l'm'} \bra{l'm'} \hat{z} \ket{lm} =
   \frac{1}{2l+1} \left[ \frac{l^2-m^2}{2l-1} \delta_{l',l-1} + \frac{(l+1)^2-m^2}{2l+3} \delta_{l',l+1} \right] \ , \\
 && \sum_{m'=-l'}^{l'} \bra{l,m\pm 1} \hat r_\pm \ket{l'm'} \bra{l'm'} \hat r_\mp \ket{l,m \pm 1} = \nl
 \hspace{20mm} = \frac{(l\pm m)(l\pm m+1)}{(2l+1)(2l-1)} \delta_{l',l-1}
               + \frac{(l\mp m)(l\mp m+1)}{(2l+1)(2l+3)} \delta_{l',l+1} \ , \\
 && \sum_{m'=-l'}^{l'} \bra{lm} \hat{z} \ket{l'm'} \bra{l'm'} \hat r_\mp \ket{l,m\pm 1} =
    \sum_{m'=-l'}^{l'} \bra{l,m\pm 1}\hat r_\pm \ket{l'm'} \bra{l'm'} \hat{z} \ket{lm} \nl
 \hspace{20mm} = \frac{c_{l, \pm m}}{2l+1}
                 \Biggsbrk{ \mp \frac{l\pm m}{2l-1} \delta_{l',l-1}
                           \pm \frac{l\mp m+1}{2l+3} \delta_{l',l+1} }  \ ,\\
 && \sum_{m'=-l'}^{l'} \bra{l,m\pm 1} \hat r_\pm \ket{l'm'} \bra{l'm'} \hat r_\pm \ket{l,m\mp 1}
  = - \frac{c_{lm} c_{l,-m}}{2l+1}
                   \Biggsbrk{ \frac{\delta_{l',l-1}}{2l-1} 
                           + \frac{\delta_{l',l+1}}{2l+3} } \ .
\ee
The coefficients of $\delta_{l',l-1}$ and $\delta_{l',l+1}$ are $c_-$ and $c_+$ respectively.
Finally, we find
\be
 \Delta E_{00}(m) \eq (- l(l+1) + 3m^2) N \ , \\
 \Delta E_{\mp \pm}(m) \eq ( l(l+1) - 3m^2 \pm (2l-1)(2l+3)m) N \ , \\
 \Delta E_{0\pm}(m\mp 1) = \Delta E_{\mp 0}(m) \eq (\mp 2l(l+1) + 3m) c_{l,\mp m} N \ , \\
 \Delta E_{++}(m-1) = \Delta E_{--}(m+1) \eq 3 c_{lm} c_{l,-m} N \ ,
\ee
where
\be
N = - \frac{\mu^3 \alpha^4}{2M m_e n^3} \frac{1}{l(l+1)(2l-1)(2l+1)(2l+3)} \ .
\ee
In the special case $l=0$, we find that
\be
\Delta E_{\pm \mp}(0) = 2 \Delta E_{00}(0) = - \frac{\mu^3 \alpha^4}{M m_e n^3} \frac{1}{3}
\ee
while $\Delta E_{\pm 0} = \Delta E_{0 \pm} = 0$.

\bibliographystyle{nb}
\bibliography{superpaper}

\end{document}